\documentclass[aps]{revtex4}
\usepackage{amsmath}
\usepackage{amssymb}
\usepackage{hyperref}
\hypersetup{colorlinks,linkcolor=red,citecolor=green,anchorcolor=black}
\begin{document}
\title{Unitarity of loop diagrams for the ghost-like $1/(k^2-M_1^2)-1/(k^2-M_2^2)$ propagator}
\author{Philip D. Mannheim}
\affiliation{Department of Physics, University of Connecticut, Storrs, CT 06269, USA\\
email: philip.mannheim@uconn.edu}
\date{January 9, 2018}

\begin{abstract}
With fourth-order derivative theories leading to propagators of the generic ghost-like $1/(k^2-M_1^2)-1/(k^2-M_2^2)$ form, it would appear that such theories have negative norm ghost states and are not unitary. However on constructing the associated quantum Hilbert space for the free theory that would produce such a propagator, Bender and Mannheim found that the Hamiltonian of the free theory is not Hermitian but is instead $PT$ symmetric, and that there are in fact no negative norm ghost states, with all Hilbert space norms being both positive and preserved in time. Even though perturbative radiative corrections cannot change the signature of a Hilbert space inner product, nonetheless it is not immediately apparent how such a ghost-like propagator would not then lead to negative probability contributions in loop diagrams. Here we obtain the relevant Feynman rules and show that all states obtained in cutting intermediate lines in loop diagrams have positive norm. Also we show that due to the specific way that unitarity (conservation of probability) is implemented in the theory, negative signatured discontinuities across cuts in loop diagrams are cancelled by a novel and unanticipated contribution of the states in which tree approximation (no loop) graphs are calculated, an effect that is foreign to standard Hermitian theories. Perturbatively, the fourth-order derivative theory is thus viable. The theory associated with the pure massless $1/k^4$ propagator is equally shown to be perturbatively viable.
\end{abstract}
\maketitle

\section{Introduction}

In a typical fourth-order derivative theory such as that based on the action
\begin{eqnarray}
I_S&=&\frac{1}{2}\int d^4x\bigg{[}\partial_{\mu}\partial_{\nu}\phi\partial^{\mu}
\partial^{\nu}\phi-(M_1^2+M_2^2)\partial_{\mu}\phi\partial^{\mu}\phi
+M_1^2M_2^2\phi^2\bigg{]},
\label{G1}
\end{eqnarray}
where $\phi(x)$ is a neutral scalar field, the equation of motion is given by
\begin{eqnarray}
&&(\partial_t^2-\nabla^2+M_1^2)(\partial_t^2-\nabla^2+M_2^2)
\phi(x)=0,
\label{G2}
\end{eqnarray}
and the phase space Hamiltonian is given by $H=\int d^3x T_{00}$, where 
\begin{eqnarray}
T_{\mu\nu}&=&\pi_{\mu}\phi_{,\nu}+\pi_{\mu}^{\phantom{\mu}\lambda}\phi_{,\nu,\lambda}-\eta_{\mu\nu}{\cal L},
\nonumber\\ 
 \pi^{\mu}&=&\frac{\partial{\cal L}}{\partial \phi_{,\mu}}-\partial_{\lambda
}\left(\frac{\partial {\cal L}}{\partial\phi_{,\mu,\lambda}}\right)=-\partial_{\lambda}\partial^{\mu}\partial^{\lambda}\phi- (M_1^2+M_2^2)\partial^{\mu}\phi,
\qquad \pi^{\mu\lambda}=\frac{\partial {\cal L}}{\partial \phi_{,\mu,\lambda}}=\partial^{\mu}\partial^{\lambda}\phi,
\nonumber\\
T_{00}&=&\pi_{0}\dot{\phi}+\frac{1}{2}\pi_{00}^2+\frac{1}{2}(M_1^2+M_2^2)\dot{
\phi}^2-\frac{1}{2}M_1^2M_2^2\phi^2
-\frac{1}{2}\pi_{ij}\pi^{ij}+\frac{1}{2}(
M_1^2+M_2^2)\phi_{,i}\phi^{,i}.
\label{G3}
\end{eqnarray}
With the use of the commutation relations given in (\ref{M81}) below the $D(x)=i\langle \Omega|T(\phi(x)\phi(0))|\Omega\rangle$ propagator obeys 
\begin{eqnarray}
&&(\partial_t^2-\nabla^2+M_1^2)(\partial_t^2-\nabla^2+M_2^2)D(x)=-\delta^4(x),
\nonumber\\
&&D(x)=\int \frac{d^4k}{(2\pi)^4}e^{-ik\cdot x}D(k)=-\int \frac{d^4k}{(2\pi)^4}\frac{e^{-ik\cdot x}}{(k^2-M_1^2)(k^2-M_2^2)},
\label{G4}
\end{eqnarray}
with partial fraction decomposition
\begin{eqnarray}
D(x)=\int \frac{d^4k}{(2\pi)^4}\frac{1}{(M_1^2-M_2^2)}\left(\frac{e^{-ik\cdot x}}{k^2-M_2^2}-\frac{e^{-ik\cdot x}}{k^2-M_1^2}\right).
\label{G5}
\end{eqnarray}
The presence of the relative minus sign given in (\ref{G5}) suggests that there would be states of negative norm in the theory, since one would anticipate identifying the propagator as $D(x)=i\langle \Omega|T(\phi(x)\phi(0))|\Omega\rangle$, with the insertion into $i\langle \Omega|T(\phi(x)\phi(0))|\Omega\rangle$ of a completeness relation for energy eigenstates of the form
\begin{eqnarray}
\sum |n\rangle\langle n|-\sum |m\rangle\langle m|=I
\label{G6}
\end{eqnarray}
immediately leading to (\ref{G5}).

However one cannot determine the structure of the underlying q-number theory purely by inspection of $D(x)$ since it itself is a c-number. Rather, one has to quantize the theory and explicitly determine the structure of the q-number Hilbert space. And when Bender and Mannheim did this, they found \cite{Bender2008a,Bender2008b} that there are in fact no states with negative norm in the theory at all. Specifically, they found that the energy eigenstates are not normalizable on the real axis, in immediate consequence of which there could not be any completeness relation of the form given in (\ref{G6}) at all since its validity  would require that the energy eigenstates be normalizable. Moreover, when acting on such non-normalizable states the Hamiltonian of the theory would not be self-adjoint since one could not throw away surface terms in an integration by parts, with the Hamiltonian thus not being Hermitian. 

In order to obtain states that are normalizable one has to continue the theory into the complex plane, and it is in the complex plane that the theory has to be formulated. And when this was done it was found that there are no states with negative norm \cite{Bender2008a,Bender2008b}, with the theory thus being viable. The key to determining the structure of the theory is thus in finding a viable set of boundary conditions, and in and of itself inspection of (\ref{G5}) does not enable one to do so.

Despite the fact the Hamiltonian is not  Hermitian, all the poles of $D(k)$ lie on the real $k_0$ axis in the complex $k_0$ plane, so all the  eigenvalues of the Hamiltonian are real. While Hermiticity of a Hamiltonian implies the reality of its eigenvalues, there is no converse theorem that says that if energy eigenvalues are real then the Hamiltonian must be Hermitian, with Hermiticity only being sufficient to yield real eigenvalues and not necessary. The necessary condition has been identified in the literature, with it being that the Hamiltonian must possess an antilinear symmetry (see e.g. \cite{Mostafazadeh2002,Solombrino2002,Mannheim2013,Mannheim2015b} and references therein), with it in addition being shown that if complex Lorentz invariance is imposed as well the antilinear symmetry is uniquely fixed to be $CPT$ \cite{Mannheim2015b,Mannheim2016b}. And in \cite{Bender2008a,Bender2008b} it was found that the Hamiltonian associated with the fourth-order derivative $I_S$ theory does in fact possess a specific antilinear symmetry, namely $PT$ ($P$ is parity and $T$ is time reversal), with $CPT$ symmetry reducing to $PT$ symmetry in this case since charge conjugation $C$ is separately conserved for a neutral scalar field theory. The $I_S$ theory thus falls into the class of non-Hermitian but $PT$-symmetric theories of the type explored by Bender and collaborators \cite{Bender1998, Bender1999,Bender2007,Special2012,Theme2013}. In these non-Hermitian but $PT$-symmetric theories the left-eigenvectors $\langle L|$ and left-vacuum $\langle \Omega_L|$ of the Hamiltonian are not the Hermitian conjugates of the right-eigenvectors $|R\rangle$ and right-vacuum $|\Omega_R\rangle$. Rather, they are related according to $\langle L|=\langle R|V$ where the operator $V$ effects $VHV^{-1}=H^{\dagger}$ \cite{Mostafazadeh2002,Mannheim2013,Mannheim2015b}, with it being the $\langle L|R\rangle$ inner product and not the $\langle R|R\rangle$ one that is time independent. In consequence, the propagator is given not by $D(x)=i\langle \Omega|T(\phi(x)\phi(0))|\Omega\rangle$, but by 
\begin{eqnarray}
D(x)=i\langle \Omega_L|T(\phi(x)\phi(0))|\Omega_R\rangle=i\langle \Omega_R|VT(\phi(x)\phi(0))|\Omega_R\rangle
\label{G7}
\end{eqnarray}
instead, with it being through the presence of the $V$ operator in (\ref{G7}) that the relative minus sign in (\ref{G5}) is generated \cite{Bender2008a,Bender2008b},  and not through any negative norm properties of the states. 

While the study of \cite{Bender2008a,Bender2008b} establishes that the Hilbert space built on $\langle \Omega_L|$ and $|\Omega_R\rangle$ possesses no states with negative norm, the theory based on $I_S$ is a non-interacting, free theory. However, if we now add on to $I_S$ an interaction term, then since one cannot change the signature of Hilbert space inner products perturbatively, radiative corrections are not able to generate any negative norm states.  Even though the theory must thus remain viable perturbatively, it is not immediately obvious how radiative loops involving the propagator  in (\ref{G5}) with its relative minus sign can actually achieve this. It is the purpose of this paper to show that radiative corrections do not lead to observable cut discontinuities in scattering amplitudes that are negative signatured, with the theory thus being viable. As we shall see, and in contrast to standard Hermitian theories, through a novel effect of the $V$ operator, negative discontinuities that occur in loop diagrams in the non-Hermitian fourth-order derivative theory are cancelled by positive discontinuities that occur in the $V$-dependent matrix elements in which tree approximation (no loop) diagrams are calculated.

\section{The Pais-Uhlenbeck two-oscillator model}

Since none of our discussion depends on the quantum-field-theoretic structure of the relativistic $I_S$ theory that is provided in \cite{Bender2008b} and in the appendix to this paper, it suffices here to study its associated quantum mechanics. Thus on setting $\omega_1=(\bar{k}^2+M_1^2)^{1/2}$, $\omega_2=(\bar{k}^2+M_2^2)^{1/2}$ and dropping the spatial dependence, the $I_S$ action reduces to the quantum-mechanical Pais-Uhlenbeck (PU) two-oscillator model action \cite{Pais1950} 
\begin{eqnarray}
I_{\rm PU}=\frac{1}{2}\int dt\left[{\ddot z}^2-\left(\omega_1^2
+\omega_2^2\right){\dot z}^2+\omega_1^2\omega_2^2z^2\right],
\label{G8}
\end{eqnarray}
where for definitiveness we take $\omega_1>\omega_2$. The equation of motion is given by
\begin{eqnarray}
\frac{d^4z}{dt^4}+(\omega_1^2+\omega_2^2)\frac{d^2z}{dt^2}+\omega_1^2\omega_2^2
z=0,
\label{G9}
\end{eqnarray}
while in analog to (\ref{G4}) and (\ref{G5}) the propagator is given by
\begin{eqnarray}
&&\left(\frac{d^4}{dt^4}+(\omega_1^2+\omega_2^2)\frac{d^2}{dt^2}+\omega_1^2\omega_2^2\right)G(t)=-\delta(t),
\nonumber\\
G(E)&=&\int dt e^{iEt}G(t)
=-\frac{1}{(E^2-\omega_1^2)(E^2-\omega_2^2)}
=\frac{1}{(\omega_1^2-\omega_2^2)}\left(\frac{1}{E^2-\omega_2^2}-\frac{1}{E^2-
\omega_1^2}\right)
\nonumber\\
&=&\frac{1}{(\omega_1^2-\omega_2^2)}\left[\frac{1}{2\omega_2}\left(\frac{1}{E-\omega_2}-\frac{1}{E+\omega_2}\right)-\frac{1}{2\omega_1}\left(\frac{1}{E-\omega_1}-\frac{1}{E+\omega_1}\right)\right].
\label{G10}
\end{eqnarray}
With $x=\dot{z}$, $[z,p_z]=i$, $[x,p_x]=i$, the Hamiltonian is given by  \cite{Mannheim2000}
\begin{eqnarray}
H_{\rm PU}=\frac{p_x^2}{2}+p_zx+\frac{1}{2}\left(\omega_1^2+\omega_2^2 \right)x^2-\frac{1}{2}\omega_1^2\omega_2^2z^2.
\label{G11}
\end{eqnarray}

With the usual causal Feynman contour prescription for the $G(E)$ propagator, positive energy states propagate forward in time, while negative energy states propagate backward in time, with the lowest positive energy eigenvalue associated with $G(E)$ being the zero-point energy of the two oscillators, viz. $E_0=(\omega_1+\omega_2)/2$. If we now set $p_z=-i\partial_z$, $p_x=-i
\partial_x$, the Schr\"odinger equation takes the form
\begin{eqnarray}
\left[-\frac{1}{2}\frac{\partial^2}{\partial x^2}-ix\frac{\partial}{
\partial z}+\frac{1}{2}(\omega_1^2+\omega_2^2)x^2-\frac{1}{2}
\omega_1^2\omega_2^2z^2\right]\psi_n(z,x)=E_n\psi_n(z,x),
\label{G12}
\end{eqnarray}
with the ground-state energy $E_0=(\omega_1+\omega_2)/2$ having eigenfunction \cite{Mannheim2007} 
\begin{eqnarray}
\psi_0(z,x)={\rm exp}\left[\frac{1}{2}(\omega_1+\omega_2)\omega_1\omega_2
z^2+i\omega_1\omega_2zx-\frac{1}{2}(\omega_1+\omega_2)x^2\right].
\label{G13}
\end{eqnarray}
As $z\to\pm\infty$, $\psi_0(z,x)$ diverges, with, as noted earlier,  the wave function not being normalizable on the real $z$ axis. 

Noting that $\psi_0(z,x)$ would be normalizable if $z$ were pure imaginary and the operator $z$ were anti-Hermitian (equivalent to representing $p_z$ by  $\partial_z$ rather than $-i\partial_z$), Bender and Mannheim continued $z$ (but not $x$) into the complex plane. However rather than work with anti-Hermitian operators,  it is instead more convenient to  make a similarity transform on the operators in $H_{PU}$ of the form \cite{Bender2008a,Bender2008b}
\begin{eqnarray}
y=e^{\pi p_zz/2}ze^{-\pi p_zz/2}=-iz,\qquad q=e^{\pi p_zz/2}p_ze^{-\pi p_zz/2}=
ip_z,
\label{G14}
\end{eqnarray}
so that $[y,q]= i$. Under this same transformation $H_{\rm PU}$ transforms into
\begin{eqnarray}
e^{\pi p_zz/2}H_{\rm PU}e^{-\pi p_zz/2}=\bar{H}=\frac{p^2}{2}-iqx+\frac{1}{2}\left(\omega_1^2+\omega_2^2
\right)x^2+\frac{1}{2}\omega_1^2\omega_2^2y^2,
\label{G15}
\end{eqnarray}
where for notational simplicity we have replaced $p_x$ by $p$, so that $[x,p]=i$. When acting on the eigenfunctions of $\bar{H}$ the  $y$ and $q$ operators are Hermitian (as are $x$ and $p$). However, as the presence of the factor $i$ in the $-iqx$ term indicates, $\bar{H}$ is not Hermitian.

\section{Quantization of the theory}

To quantize the theory one sets \cite{Bender2008b}
\begin{eqnarray}
\dot{y}(t)&=&i[\bar{H},y]=-ix(t),\qquad \dot{x}(t)=p(t),\qquad
\dot{p}(t)=iq(t)-(\omega_1^2+\omega_2^2)x(t),\qquad
\dot{q}(t)=-\omega_1^2\omega_2^2y(t),
\nonumber\\
y(t)&=&-ia_1e^{-i\omega_1t}+a_2e^{-i\omega_2t}-i\hat{a}_1e^{i\omega_1t}+\hat{a}_2e^{i\omega_2t},
\nonumber\\
x(t)&=&-i\omega_1a_1e^{-i\omega_1t}+\omega_2a_2e^{-i\omega_2t}+i\omega_1\hat{a}_1
e^{i\omega_1t}-\omega_2\hat{a}_2e^{i\omega_2t},
\nonumber\\
p(t)&=&-\omega_1^2a_1e^{-i\omega_1t}-i\omega_2^2a_2e^{-i\omega_2t}-\omega_1
^2\hat{a}_1e^{i\omega_1t}-i\omega_2^2\hat{a}_2e^{i\omega_2t},
\nonumber\\
q(t)&=&\omega_1\omega_2[-\omega_2a_1e^{-i\omega_1t}-i\omega_1a_2e^{-i
\omega_2t}+\omega_2\hat{a}_1e^{i\omega_1t}+i\omega_1\hat{a}_2e^{i\omega_2t}],
\nonumber\\
a_1e^{-i\omega_1t}&=&\frac{1}{2(\omega_1^2-\omega_2^2)}\left[-i\omega_2^2y-p+i\omega_1x+\frac{q}{\omega_1}\right],\qquad \hat{a}_1e^{+i\omega_1t}=\frac{1}{2(\omega_1^2-\omega_2^2)}\left[-i\omega_2^2y-p-i\omega_1x-\frac{q}{\omega_1}\right],
\nonumber\\
a_2e^{-i\omega_2t}&=&\frac{1}{2(\omega_1^2-\omega_2^2)}\left[\omega_1^2y-ip-\omega_2x+\frac{iq}{\omega_2}\right],\qquad \hat{a}_2e^{+i\omega_2t}=\frac{1}{2(\omega_1^2-\omega_2^2)}\left[\omega_1^2y-ip+\omega_2x-\frac{iq}{\omega_2}\right].
\label{G16}
\end{eqnarray}
With $[x,p]=i$, $[y,q]=i$, the $a_i$ and $\hat{a}_i$ operators obey the standard two-oscillator commutation algebra 
\begin{eqnarray}
&&[a_1,\hat{a}_1]=\frac{1}{2\omega_1(\omega_1^2-\omega_2^2)}=N_1,~~
[a_2,\hat{a}_2]=\frac{1}{2\omega_2(\omega_1^2-\omega_2^2)}=N_2,
\nonumber\\
&&[a_1,a_2]=0,\qquad [a_1,\hat{a}_2]=0,\qquad [\hat{a}_1,a_2]=0,\quad [\hat{a}_1,\hat{a}_2]=0,
\label{G17}
\end{eqnarray}
with $\hat{a}_1$ and $\hat{a}_2$ serving as creation operators and $a_1$ and $a_2$ serving as annihilators in the Hilbert space built on the right- and left-vacua $|\Omega_R\rangle$ and $\langle \Omega_L|$ according to 
\begin{eqnarray}
a_1|\Omega_R\rangle=0,\qquad a_2|\Omega_R\rangle=0,\qquad \langle \Omega_L|\hat{a}_1=0, \qquad \langle \Omega_L|\hat{a}_2=0. 
\label{G18}
\end{eqnarray}
With both $N_1$ and $N_2$ being positive there are no states with negative norm. Because the theory is not Hermitian the $\hat{a}_i$ are not the Hermitian conjugates of the $a_i$, and $\langle \Omega_L|$ is not the Hermitian conjugate of $|\Omega_R\rangle$. However, since the theory is a $PT$ symmetric one, one can characterize the  $x$, $y$, $p$ and $q$ operators by their behavior under $PT$, with $x$ and $y$ being taken to be $PT$ odd and $p$ and $q$ to be $PT$ even in \cite{Bender2008a,Bender2008b}. In consequence $a_1$ and $\hat{a}_1$ are $PT$ even and  $a_2$ and $\hat{a}_2$ are $PT$ odd. We should note that these $PT$ assignments are not unique, and in \cite{Mannheim2017} $x$ and $y$ were taken to be $PT$ even and $p$ and $q$ to be $PT$ odd, with $a_1$ and $\hat{a}_1$ then being $PT$ odd and  $a_2$ and $\hat{a}_2$ being $PT$ even.  Both of these sets of assignments are consistent with (\ref{G16}) and the $PT$ symmetry of $\bar{H}$, and none of these choices affect the conclusions of this paper.

In terms of the creation and annihilation operators the Hamiltonian is diagonalized as
\begin{eqnarray}
\bar{H}=2(\omega_1^2-\omega_2^2)[\omega_1^2\hat{a}_1a_1+\omega_2^2\hat{a}_2a_2]
+\frac{1}{2}(\omega_1+\omega_2),
\label{G19}
\end{eqnarray}
and with there being no relative minus sign between the $\omega_1^2\hat{a}_1a_1$ and $\omega_2^2\hat{a}_2a_2$ terms, there are no states of negative energy. On dropping the zero point energy for convenience, for this Hamiltonian the right- and left-vacua obey $\bar{H}|\Omega_R\rangle=0$, $\langle \Omega_L|\bar{H}=0$.

The one-particle right-eigenvector states that obey $\bar{H}|i\rangle=E_i|i\rangle=\omega_i|i\rangle$ are  
\begin{eqnarray}
|1_R\rangle=N_1^{-1/2}\hat{a}_1|\Omega_R\rangle,~~|2_R\rangle=N_2^{-1/2}\hat{a}_2|\Omega_R\rangle,
\label{G20}
\end{eqnarray}
while the left-eigenvector one-particle states that obey $\langle i|\bar{H}=\langle i|E_i=\langle i|\omega_i$ are
\begin{eqnarray}
\langle 1_L|=N_1^{-1/2}\langle \Omega_L| a_1 ,~~\langle 2_L|=N_2^{-1/2}\langle \Omega_L| a_2.
\label{G21}
\end{eqnarray}
These states obey the  orthonormal relations
\begin{eqnarray}
\langle 1_L|1_R\rangle=1,\qquad\langle 1_L|2_R\rangle=0,\qquad\langle 2_L|1_R\rangle=0,\qquad\langle 2_L|2_R\rangle=1,\qquad |1_R\rangle\langle 1_L|+|2_R\rangle\langle 2_L|=I,
\label{G22}
\end{eqnarray}
to thus have positive norm and obey a positive signatured closure relation. 

Because we have replaced $z$ by $-iz=y$, $p_z$ by $ip_z=q$, and $H_{\rm PU}$ by $\bar{H}$, the propagator is now given by $\bar{G}(t)=-i\langle \Omega_L|y(t)y(0)|\Omega_R\rangle$, (equivalent to $+i\langle \Omega_L|z(t)z(0)|\Omega_R\rangle$ rather than $-i\langle \Omega_L|z(t)z(0)|\Omega_R\rangle$), and it  obeys 
\begin{eqnarray}
&&\left(\frac{d^4}{dt^4}+(\omega_1^2+\omega_2^2)\frac{d^2}{dt^2}+\omega_1^2\omega_2^2\right)\bar{G}(t)=-\delta(t),
\nonumber\\
\bar{G}(E)&=&-\frac{1}{(E^2-\omega_1^2)(E^2-\omega_2^2)}=\frac{1}{\omega_1^2-\omega_2^2}\left(-\frac{1}{E^2-\omega_1^2}+\frac{1}{E^2-
\omega_2^2}\right)
\nonumber\\
&=&\frac{1}{(\omega_1^2-\omega_2^2)}\left[-\frac{1}{2\omega_1}\left(\frac{1}{E-\omega_1}-\frac{1}{E+\omega_1}\right)+\frac{1}{2\omega_2}\left(\frac{1}{E-\omega_2}-\frac{1}{E+\omega_2}\right)\right].
\label{G23}
\end{eqnarray}
Given the form for $y(t)$ given in (\ref{G16}) we evaluate the two-point function, and for $t>0$ obtain 
\begin{eqnarray}
\bar{G}(t)&=&-i\langle \Omega_L|y(t)y(0)|\Omega_R\rangle
\nonumber\\
&=&-i\langle \Omega_L|[-ia_1e^{-i\omega_1t}+a_2e^{-i\omega_2t}][-i\hat{a}_1+\hat{a}_2]|\Omega_R\rangle
\nonumber\\
&=&-i[-N_1e^{-i\omega_1 t}+N_2e^{-i\omega_2 t}].
\label{G24}
\end{eqnarray}
In Fourier space $\bar{G}(E)=\int dt e^{iEt}\theta(t)\bar{G}(t)$ is given by 
\begin{eqnarray}
\bar{G}(E)=-\frac{N_1}{E-\omega_1+i\epsilon}+\frac{N_2}{E-\omega_2+i\epsilon},
\label{G25}
\end{eqnarray}
with the minus sign in the $\omega_1$ term being generated by the $-i$ factors in $y(t)$ and not by states of negative norm. Comparing the positive frequency components of (\ref{G23}) with (\ref{G25}) we confirm that despite our being in a fourth-order derivative theory rather than a second-order one, the propagator is given by $\bar{G}(t)=-i\langle \Omega_L|y(t)y(0)|\Omega_R\rangle$ and not by any expression that might involve time derivatives of $y(t)$. Consequently, in the Feynman rules for the interacting theory that we discuss below, the Wick contraction procedure will be the completely standard second-order theory one familiar from Hermitian theories.

This expression that we have obtained for $\bar{G}(E)$ is just the value of $G(E)$ given in (\ref{G10}) as evaluated at the two positive frequency poles $\omega=\omega_1$ and $\omega=\omega_2$. (We defined the propagators so that the coefficient of the $\delta(t)$ term would be negative in both (\ref{G10}) and (\ref{G23})). In (\ref{G25}) the $\omega_2$ term has the same overall positive sign as that of a standard one-dimensional harmonic oscillator, while the $\omega_1$ term has the opposite  sign to that of  a standard one-dimensional harmonic oscillator. (In a standard second-order field theory where $\delta(t)[\dot{\phi}(x),\phi(0)]=-i\delta^4(x)$, we have $(\partial_t^2-\nabla^2+m^2)[-i\langle \Omega|T(\phi(x)\phi(0))|\Omega\rangle]=(\partial_t^2-\nabla^2+m^2)D(x)=-\delta^4(x)$ (viz. a negative coefficient), so that $D(x)=+\int d^4k/(2\pi)^4e^{-ik\cdot x}/(k^2-m^2)$.) As we see, we can obtain a minus sign in $\bar{G}(E)$ even though no commutator in (\ref{G17}) is negative. A relative minus sign in the propagator is thus not indicative of the presence of states with negative norm, and in this way the theory is viable.  This same analysis goes through identically in the relativistic case of $I_S$ itself, and we present the calculation in an appendix.

To underscore that all norms are positive we note that we can make a similarity transformation on $\bar{H}$ in order to decouple the two oscillators \cite{Bender2008a,Bender2008b}. Specifically, one introduces an operator $Q$
\begin{eqnarray}
Q=\alpha pq+\beta xy,\qquad \alpha=\frac{1}{\omega_1\omega_2}{\rm log}\left(\frac{\omega_1+\omega_2}{\omega_1-\omega_2}\right),\qquad \beta=\alpha\omega_1^2\omega_2^2,
\label{G26}
\end{eqnarray}
with $Q$ being Hermitian since $x$, $y$, $p$ and $q$  are all Hermitian, while being $PT$ even for either choice of $PT$ assignments of the $x$, $y$, $p$ and $q$  operators  described above. With this $Q$ we transform $\bar{H}$ of (\ref{G15}) and $x$, $y$, $p$ and $q$ according to
\begin{eqnarray}
\bar{H}^{\prime}&=&e^{-Q/2}\bar{H}e^{Q/2}=\frac{p^{\prime 2}}{2}-iq^{\prime}x^{\prime}+\frac{1}{2}\left(\omega_1^2+\omega_2^2
\right)x^{\prime 2}+\frac{1}{2}\omega_1^2\omega_2^2y^{\prime 2},
\nonumber\\
y^{\prime}&=&e^{-Q/2}ye^{Q/2}=y\cosh\theta+i(\alpha/\beta)^{1/2}p\sinh\theta, \qquad
p^{\prime}=e^{-Q/2}ye^{Q/2}=p\cosh\theta-i(\beta/\alpha)^{1/2}y\sinh\theta,
\nonumber\\
x^{\prime}&=&e^{-Q/2}xe^{Q/2}=x\cosh\theta+i(\alpha/\beta)^{1/2}q\sinh\theta, \qquad
q^{\prime}=e^{-Q/2}ye^{Q/2}=q\cosh\theta-i(\beta/\alpha)^{1/2}x\sinh\theta,
\label{N27}
\end{eqnarray}
where $\theta=(\alpha\beta)^{1/2}/2$, $\tanh\theta=\omega_2/\omega_1$. However,  as noted in  \cite{Bender2008a} the transformed $\bar{H}^{\prime}$ can actually be written entirely in terms of the original $x$, $y$, $p$ and $q$ variables as 
\begin{eqnarray}
&&\bar{H}^{\prime}
=\frac{p^2}{2}+\frac{q^2}{2\omega_1^2}+
\frac{1}{2}\omega_1^2x^2+\frac{1}{2}\omega_1^2\omega_2^2y^2.
\label{N28}
\end{eqnarray}
In addition we note that with its phase being $-Q/2$ rather than $-iQ/2$, the  $e^{-Q/2}$ operator is not unitary. As we see from (\ref{N28}), with this non-unitary transformation we bring $\bar{H}$ to a Hermitian form, since with $x$, $y$, $p$ and $q$ all being Hermitian, it follows that $\bar{H}^{\prime}$ is Hermitian too. With $\bar{H}^{\prime}$ having the form of two decoupled, conventional, ghost-free harmonic oscillators each with positive norm states and positive energy modes, we confirm, just as we had found, that the same must be true of the untransformed $\bar{H}$ since one cannot change the signs of inner products or energies by a similarity transformation. 

In addition we note that $e^{-Q}$ effects $e^{-Q}\bar{H}e^{Q}=\bar{H}^{\dagger}$, a relation that entails that $\bar{H}$ and $\bar{H}^{\dagger}$ are isospectrally related. $\bar{H}$ and $\bar{H}^{\dagger}$ thus have to have the same set of eigenvalues, consistent with the fact that all eigenvalues of $\bar{H}$ are real. For the PU oscillator we recognize $e^{-Q}$ as being the operator $V$ that we introduced earlier, and we can thus write the PU oscillator propagator as 
\begin{eqnarray}
D(t)=-i\langle \Omega_L|T(y(t)y(0))|\Omega_R\rangle=-i\langle \Omega_R|e^{-Q}T(y(t)y(0))|\Omega_R\rangle.
\label{N29}
\end{eqnarray}

\section{Radiative corrections and loop diagrams}

In analyzing loop diagrams with a total incoming momentum $P_{\mu}$ we have to consider two situations, $P_{\mu}$ timelike and $P_{\mu}$ spacelike. For  spacelike $P_{\mu}$ the poles in loop diagrams with a running momentum $k_{\mu}$ lie in the upper-left or lower-right quadrants in the complex $k_0$ plane. The Feynman contour  encloses the lower-left and lower-right quadrants, and for spacelike $P_{\mu}$ thus encloses those poles with positive frequency. The Wick  contour encloses the lower-left and upper-right quadrants and thus encloses no poles  at all, and pole structure has no bearing on the evaluation of the spacelike region Green's functions. The Wick contour can be used to explore the off-shell spacelike  Euclidean behavior in $P_{\mu}P^{\mu}$ of loop diagrams, and to determine the behavior in the timelike region the resulting expressions can then be analytically continued in $P_{\mu}P^{\mu}$ in order to identify on-shell branch points and branch cuts. Alternatively, in the timelike region one could evaluate the Feynman contour directly. Since momenta are far off-shell in the asymptotic spacelike region, in that region there is no concern about the relative minus sign in the propagator affecting probabilities (an on-shell concept), and one can take advantage of the $1/k^4$ convergence property of the asymptotic spacelike propagator. However, we do need to address how the relative minus sign in the propagator might affect the on-shell structure of loop diagrams in the timelike region. 

\subsection{The Feynman rules}

In order to evaluate intermediate loops we need to determine the appropriate Feynman rules. We note that the fourth-order propagator only depends on the field and not any of its time derivatives, i.e. $-i\langle \Omega_L|T(y(t)y(0))|\Omega_R\rangle$ in the PU case and $-i\langle \Omega_L|T(\bar{\phi}(x)\bar{\phi}(0))|\Omega_R\rangle$ in the relativistic case where $\bar{\phi}=-i\phi$, which as discussed below, is the analog of $y=-iz$. Since it is immaterial to the Wick contraction procedure as to whether or not $\langle \Omega_L|$ is the Hermitian conjugate of $|\Omega_R\rangle$, the contraction rules for a string of field operators are exactly the same as in the standard second-order Hermitian case. Similarly, the rules for vertices are the same as well. Thus for a  $-\lambda y^4=-\lambda z^4$ or a $-\lambda \bar{\phi}^4=-\lambda \phi^4$ interaction for instance, (each of these interactions being $PT$ symmetric no matter whether $y$, $z$, $\bar{\phi}$ or $\phi$ are $PT$ even or $PT$ odd), in a two-particle to two-particle scattering amplitude the one loop graphs (shaped like the symbol $\between$) would contain two intermediate lines and respectively behave as 
\begin{eqnarray}
\Pi(E)&=&i\lambda^2\int \frac{d\omega}{2\pi}\frac{1}{(\omega^2-\omega_1^2)(\omega^2-\omega_2^2)}\frac{1}{((\omega+E)^2-\omega_1^2)((\omega+E)^2-\omega_2^2)}, 
\nonumber\\
\Pi(P^{\mu}P_{\mu})&=&i\lambda^2\int \frac{d^4k}{(2\pi)^4}\frac{1}{(k^2-M_1^2)(k^2-M_2^2)}\frac{1}{((k+p)^2-M_1^2)((k+p)^2-M_2^2)}
\nonumber\\ 
&=&\frac{i\lambda^2}{(M_1^2-M_2^2)^2}\int d^4k\left(\frac{1}{k^2-M_2^2}-\frac{1}{k^2-M_1^2}\right)
\left(\frac{1}{(k+p)^2-M_2^2}-\frac{1}{(k+p)^2-M_1^2}\right).
\label{N30}
\end{eqnarray}
In (\ref{N30}) we have used $\bar{G}(E)$ and the $\bar{D}(x)=-i\langle \Omega_L|T(\bar{\phi}(x)\bar{\phi}(0))|\Omega_R\rangle$ propagator given in (\ref{M82}) below. Thus the Wick contraction procedure is the standard one, with the only difference being that one should use the fourth-order derivative theory propagator rather than the standard second-order one, with there being no modification in the rules for vertices.

\subsection{The cutting rules}

Suppose we now try to cut propagators in intermediate loops. We can introduce on-shell intermediate $|1_R\rangle, |2_R\rangle, \langle 1_L|, \langle 2_L|$ states and evaluate their contribution to the $\bar{G}(t)=-i\langle \Omega_L|y(t)y(0)|\Omega_R\rangle$ propagator with $t>0$, to obtain
\begin{eqnarray}
\bar{G}(t)&=&-i\langle \Omega_L|y(t)y(0)|\Omega_R\rangle
\nonumber\\
&=&-i\langle \Omega_L|y(t)|1_R\rangle \langle 1_L|y(0)|\Omega_R\rangle
-i\langle \Omega_L|y(t)|2_R\rangle \langle 2_L|y(0)|\Omega_R\rangle
\nonumber\\
&=&-i\langle \Omega_L|(-ia_1e^{-i\omega_1t})|1_R\rangle \langle 1_L|(-i\hat{a}_1)|\Omega_R\rangle
-i\langle \Omega_L|a_2e^{-i\omega_2t}|2_R\rangle \langle 2_L|\hat{a}_2|\Omega_R\rangle
\nonumber\\
&=&-i[-N_1e^{-i\omega_1 t}+N_2e^{-i\omega_2 t}].
\label{N31}
\end{eqnarray}
Recognizing (\ref{N31}) as being none other than (\ref{G24}), we thus see that the intermediate on-shell states associated with cutting the $\bar{G}(t)$ propagator all have positive norm. Thus despite the relative minus sign in $\bar{G}(E)$ (or analogously in $\bar{D}(k)=\int d^4xe^{ik\cdot x}\theta(t)\bar{D}(x)$, where we cut using the states introduced in the appendix), all cut lines have positivity, just as required of the standard Landau-Cutkosky cutting rules As we shall see below, in the field theory case these cutting rules are consistent with the existence of on-shell branch cuts in scattering amplitudes, with the masses of the intermediate on-shell states fixing the locations of branch points.

\subsection{Pseudo-unitarity and the time evolution operator}

In order to be as general as possible in discussing unitarity let us consider some general time-independent Hamiltonian $H$ that does not obey $H=H^{\dagger}$ (as usual the dagger symbol denotes the operation complex conjugation plus transposition), but for the moment let us put no other constraint on it. Let us introduce right-eigenvector states of $H$ that obey
\begin{eqnarray}
i\partial_t|R_n\rangle=H|R_n\rangle,\qquad -i\partial_t\langle R_n|=\langle R_n|H^{\dagger}.
\label{N32}
\end{eqnarray}
Because $H$ is not Hermitian, we see that $\langle R_n|$ is not a left-eigenvector of $H$, with the matrix element $\langle R_n(t)|R_n(t)\rangle=\langle R_n(0)|e^{iH^{\dagger}t} e^{-iHt}|R_n(0)\rangle$ not being equal to $\langle R_n(0)|R_n(0)\rangle$, to thus not be time independent. To construct a Hilbert space inner product that is time independent, we introduce a time-independent operator $V$ and a $V$ norm $\langle R_n|V|R_m\rangle$ that thus obeys
\begin{eqnarray}
i\partial_t\langle R_n|V|R_m\rangle=\langle R_n|(VH-H^{\dagger}V)|R_m\rangle.
\label{N33}
\end{eqnarray}
Then if $V$ obeys
\begin{eqnarray}
VH-H^{\dagger}V=0,
\label{N34}
\end{eqnarray}
the $V$ norm will be preserved in time, with it being the $V$ norm rather than the standard $\langle R_n|R_m\rangle$ Dirac norm that is the one that is needed in the non-Hermitian case. In \cite{Mannheim2015b} a converse theorem was established, namely that if we start with the time independence of the $V$ norm as input, then if the $|R_m\rangle$ states are complete we can establish the relation $VH-H^{\dagger}V=0$ as an operator identity.  The validity of the relation $VH=H^{\dagger}V$ entails the time independence of inner products, while the time independence of inner products entails that $VH-H^{\dagger}V=0$. Probability conservation in the sense of the time independence of inner products  and probability conservation in the sense of completeness of states are thus just as intimately connected in the non-Hermitian case as they are in the Hermitian case, and the interplay of these two notions of probability conservation will be central to the analysis of this paper.  

For a $V$ that is  invertible (this for instance being the case for the $V=e^{-Q}$ operator in the PU theory or its relativistic generalization), we in addition obtain the so-called pseudo-Hermitian condition \cite{Mostafazadeh2002,Solombrino2002}
\begin{eqnarray}
VHV^{-1}=H^{\dagger}.
\label{N35}
\end{eqnarray}
With $H$ and $H^{\dagger}$ thus being isospectrally related when $V$ is invertible, they must have the same set of eigenvalues, just as required of the PU oscillator theory where all eigenvalues are real. Moreover, as noted in general in \cite{Mannheim2015b}, any Hamiltonian that obeys (\ref{N35}) must necessarily possess an antilinear symmetry. Requiring that inner products be time independent is thus equivalent to demanding antilinear symmetry. If one in addition imposes complex Lorentz invariance, the antilinear symmetry is uniquely specified to be $CPT$ \cite{Mannheim2015b,Mannheim2016b}. With the neutral scalar field $I_S$ action being both $CPT$ and $C$ invariant, the $PT$ symmetry of the PU oscillator theory that descends from it is thus automatic.  

Given (\ref{N35}), we can construct states that are left-eigenvectors of $H$. Specifically, we note that
\begin{eqnarray}
 -i\partial_t\langle R_n|V=\langle R_n|H^{\dagger}V=\langle R_n|VH.
\label{N36}
\end{eqnarray}
We can thus identify the left-eigenvectors as $\langle L_n|=\langle R_n|V$ (here $n$ labels the eigenvector and not one of its components in any chosen basis) as they obey $-i\partial_t\langle L_n|=\langle L_n|H.$ And since they do, the so-called biorthogonal inner product  $\langle L_n(t)|R_n(t)\rangle=\langle L_n(0)|e^{iHt} e^{-iHt}|R_n(0)\rangle=\langle L_n(0)|R_n(0)\rangle$ is preserved in time.

In the same way that the operator $V$ generates pseudo-Hermiticity it also generates pseudo-unitarity, with the time evolution operator $U=e^{-iHt}$ obeying 
\begin{eqnarray}
UV^{-1}U^{\dagger}V=UV^{-1}e^{iH^{\dagger}t}V=Ue^{iHt}=UU^{-1}=I.
\label{N37}
\end{eqnarray}
It is by obeying the relation $UV^{-1}U^{\dagger}V=I$ that $U$ is able to evolve states so that their norms remain unchanged in time.

With the $S$ matrix being the double limit $t_i\rightarrow -\infty$, $t_f\rightarrow \infty$ of $U=e^{-iH(t_f-t_i)}$, the $S$ matrix obeys
\begin{eqnarray}
SV^{-1}S^{\dagger}V=I,
\label{N38}
\end{eqnarray}
a relation that reduces to the familiar $SS^{\dagger}=I$ when $V=I$ and $H=H^{\dagger}$. Since $S$ is the limit of a time evolution operator $U$ that maintains the time independence of the left-right inner product, scattering must also maintain probability. Given such an assurance, scattering involving the $1/(k^2-M_2^2)-1/(k^2-M_1^2)$ propagator must thus, despite its appearance, preserve probability too. How it actually does so will be shown in more detail below by introducing the $T$ matrix.

Now instead of just considering Hamiltonians that obey $H^{\dagger}=VHV^{-1}$, let us in addition restrict to Hamiltonians that have all eigenvalues real and all eigenvectors complete. (As well as allowing for eigenvalues to be real the isospectral equivalence of $H$ and $H^{\dagger}$ also allows for eigenvalues to appear in complex conjugate pairs.) Such Hamiltonians must either already obey $H=H^{\dagger}$ or be transformable by a (non-unitary) similarity transformation $B$ into one that does according to $BHB^{-1}=H^{\prime}=H^{\prime \dagger}$. (With $H$ being taken to be time independent here, $B$ can be taken to be time independent too.) For the primed system one has eigenvectors that obey 
\begin{eqnarray}
 i\partial_t|R_n^{\prime}\rangle=H^{\prime}|R_n^{\prime}\rangle,\qquad -i\partial_t\langle R_n^{\prime}|=\langle R_n^{\prime}|H^{\prime},
\label{N39}
\end{eqnarray}
with the eigenstates of $H$ and $H^{\prime}$ being related by
\begin{eqnarray}
 |R_n^{\prime}\rangle=B|R_n\rangle,~~~\langle R_n^{\prime}|=\langle R_n|B^{\dagger}.
\label{N40}
\end{eqnarray}
On normalizing the eigenstates of $H^{\prime}$ to unity,  we obtain
\begin{eqnarray}
\langle R_n^{\prime} |R_m^{\prime}\rangle=\langle R_n|B^{\dagger}B|R_m\rangle=\delta_{m,n}.
\label{N41}
\end{eqnarray}
With $H^{\prime}=H^{\prime \dagger}$ we obtain
\begin{eqnarray}
BHB^{-1}=B^{\dagger -1}H^{\dagger}B^{\dagger},\qquad B^{\dagger}BHB^{-1}B^{\dagger -1}=B^{\dagger}BH[B^{\dagger }B]^{-1}=H^{\dagger}.
\label{N42}
\end{eqnarray}
We can thus identify $B^{\dagger}B$ with $V$, and as noted in \cite{Mannheim2017}, can thus establish that the $V$ norm is the $B^{\dagger}B$ norm, so that in this case $\langle R_n|V|R_m\rangle=\langle L_n|R_m\rangle=\delta_{m,n}$ is positive definite. The interpretation of the $V$ norms as probabilities is then secured, with their time independence ensuring that probability is preserved in time.

\subsection{Pseudo-unitarity and in and out states}

While we can establish the pseudo-unitarity of the $S$ matrix by relating it to the evolution operator, because we are in a non-Hermitian theory it turns out that we cannot also establish the same pseudo-unitarity condition by using the completeness properties of in and out states. Using the $S$ matrix to relate in and out states actually makes no reference to the interactions that might occur between states coming in at $t_i=-\infty$ and  going out at $t_f=+\infty$ (these interactions being taken to be adiabatically switched on at $t_i=-\infty$ and switched off at $t_f=+\infty$), and thus only involves solutions to the non-interacting Hamiltonian, i.e. it only involves the eigenstates of Hamiltonians such as the free $H_{\rm PU}$. The in-out definition of the S matrix does not involve the operator $V$ that acts while the interaction is active, and thus cannot recover (\ref{N38}) unless we are in a Hermitian theory and $V=I$. Rather, the in-out definition can only involve the $V_{in}$ and $V_{out}$ operators associated with the in and out states themselves. Thus all that is needed for the in-out definition is the completeness of the in states and completeness of the out states. For free in and out theories this is the case as long as the free Hamiltonian is not of non-diagonalizable Jordan-block form, a form that we will actually encounter below when we discuss the equal-frequency PU oscillator theory. For the PU case $H_{\rm PU}$ is similarity equivalent to the Hermitian $\bar{H}^{\prime}$ (cf. (\ref{N28})), so the eigenstates of $H_{\rm PU}$ are indeed complete. In fact, completeness of the eigenstates of $H_{\rm PU}$ is not in question since in writing the propagator as $1/(E^2-\omega_1^2)-1/(E^2-\omega_2^2)$ (or as $1/(k^2-M_1^2)-1/(k^2-M_2^2)$) we are writing the propagator as the sum of two standard propagators both of whose eigenspectra are complete. The relative minus sign does not affect completeness, but as noted in \cite{Bender2008a,Bender2008b} and described above, it does affect what constitutes the appropriate set of eigenvectors.

While we shall take $V_{in}$ and $V_{out}$ to be Hermitian, we note that for the in and out states of any free Hamiltonian (of which  $H_{\rm PU}$ is an example), the relevant $V_{in}$ and $V_{out}$ need not actually be Hermitian, since in the event that the Hamiltonian has complex conjugate eigenvalues (this would actually be the case for $H_{\rm PU}$ if one makes the $PT$-symmetry-preserving substitution $\omega_1=\omega_R+i\omega_I$, $\omega_2=\omega_R-i\omega_I$ \cite{Mannheim2015b}), complex conjugate pairs of eigenvectors would be needed for completeness and the relevant $V$ would not be Hermitian. (A particle such as a pion is a stable out state under strong interaction scattering but not under electroweak scattering.) 

On restricting  to Hermitian $V_{in}$ and $V_{out}$, the left and right in and out eigenvectors are related by $\langle L_{in}|=\langle R_{in}|V_{in}$ and $V_{in}|R_{in}\rangle=|L_{in}\rangle$ and $\langle L_{out}|=\langle R_{out}|V_{out}$ and $V_{out}|R_{out}\rangle=|L_{out}\rangle$. They obey the biorthonormal conditions  (see e.g. \cite{Mannheim2013})
\begin{eqnarray}
&&\langle L_{in}^i|R_{in}^j\rangle=\langle R_{in}^i|V_{in}|R_{in}^j\rangle=\delta_{i,j},\qquad \langle L_{out}^i|R_{out}^j\rangle=\langle R_{out}^i|V_{out}|R_{out}^j\rangle=\delta_{i,j},
\nonumber\\
&&\sum_i|R^i_{in}\rangle\langle R_{in}^i|V_{in}=
\sum_iV_{in}|R^i_{in}\rangle\langle R_{in}^i|=I,~~
\sum_i|R^i_{out}\rangle\langle R_{out}^i|V_{out}= \sum_iV_{out}|R^i_{out}\rangle\langle R_{out}^i|=I.
\label{N43}
\end{eqnarray}
In terms of these in and out states the $S$ matrix operator that effects $\langle L_{in}^i|S=\langle L_{out}^i|$, $|L_{out}^i\rangle=S^{\dagger}|L_{in}^i\rangle$, $|R_{out}^i\rangle=V_{out}^{-1}S^{\dagger}V_{in}|R_{in}^i\rangle$ is given by
\begin{eqnarray}
S=\sum_{i}|R_{in}^i\rangle\langle L_{out}^i|=\sum_{i}|R_{in}^i\rangle\langle R_{out}^i|V_{out},
\qquad S^{\dagger}=\sum_{i}|L_{out}^i\rangle\langle R_{in}^i|=\sum_{i}V_{out}|R_{out}^i\rangle\langle R_{in}^i|.
\label{N44}
\end{eqnarray}
Thus from (\ref{N43}) and (\ref{N44}) we obtain
\begin{equation}
SV_{out}^{-1}S^{\dagger}V_{in}=\sum_i|R_{in}^i\rangle\langle R_{out}^i|V_{out}V_{out}^{-1}\sum_jV_{out}|R_{out}^j\rangle\langle R_{in}^j|V_{in}=\sum_i\sum_j|R_{in}^i\rangle\delta_{i,j}\langle R_{in}^j|V_{in}=\sum_i|R_{in}^i\rangle\langle R_{in}^i|V_{in}=I.
\label{N45}
\end{equation}
Consequently, for non-Hermitian but $PT$- or $CPT$-invariant Hamiltonians in-out pseudo-unitarity of the scattering process takes the form 
\begin{eqnarray}
SV_{out}^{-1}S^{\dagger}V_{in}=I. 
\label{N46}
\end{eqnarray}
We note that (\ref{N46}) differs from (\ref{N38}). While (\ref{N46})  is a valid relation,  in the following we shall use (\ref{N38}) as it directly connects to the time evolution operator and its Feynman diagram expansion. Even if a Hamiltonian is not Hermitian it is still the generator of time translations, and thus in the following we shall understand the $S$ matrix to be the asymptotic time limit of $U(t_f,t_i)$. (Even for potential scattering time-evolution-operator unitarity differs from in-out unitarity when a Hamiltonian is not Hermitian, and it is time evolution unitarity that is relevant. This point will be discussed in detail elsewhere \cite{Brandstotter2018}.)

\subsection{Unitarity and CPT}

Another way to implement a unitarity condition is to note \cite{Mannheim2013} that when $H$ is $CPT$ invariant  one can set $CPTe^{-iHt}[CPT]^{-1}=e^{iHt}=U^{-1}$. Thus $UCPTU[CPT]^{-1}=I$.

\subsection{The T matrix and the signs of discontinuities across cuts}

To connect to specific scattering diagrams we introduce the $T$ matrix according to
\begin{eqnarray}
S=I-iT,\qquad S^{\dagger}=I+iT^{\dagger}.
\label{N47}
\end{eqnarray}
For convenience, both here in (\ref{N47}) and throughout we  suppress a four-momentum conserving delta function in the definition of $T$. From the pseudo-unitarity condition $SV^{-1}S^{\dagger}V=I$ given in (\ref{N38}) we obtain
\begin{eqnarray}
T-V^{-1}T^{\dagger}V=-iTV^{-1}T^{\dagger}V.
\label{N48}
\end{eqnarray}
Now while we were able to derive (\ref{N38}) without needing to require that $V$ be Hermitian, in order to establish the biorthonormality of the basis vectors, which we will need below, we need $V$ to be writable as the Hermitian $V=B^{\dagger}B$ where $B$ implements $BHB^{-1}=H^{\prime}=H^{\prime \dagger}$, with the theory being similarity equivalent to a Hermitian theory. We shall thus restrict to such $V$ operators, since for them we have both biorthonormality and closure relations of the form  
\begin{eqnarray}
\langle R_n|V|R_m\rangle=\langle L_n|R_m\rangle=\delta_{m,n},\qquad 
|R_n\rangle\langle R_n|V=|R_n\rangle\langle L_n|=I.
\label{M49}
\end{eqnarray}

On taking left-right matrix elements of the left-hand side of (\ref{N48}) we obtain
\begin{eqnarray}
\langle R_{\alpha}|V(T-V^{-1}T^{\dagger}V)|R_{\beta}\rangle=\langle R_{\alpha}|VT|R_{\beta}\rangle-\langle R_{\alpha}|T^{\dagger}V|R_{\beta}\rangle
=\langle R_{\alpha}|VT|R_{\beta}\rangle-\langle R_{\beta}|VT|R_{\alpha}\rangle^*.
\label{M50}
\end{eqnarray}
Similarly, from the right-hand side of (\ref{N48}), through use of the closure relation we obtain
\begin{eqnarray}
\langle R_{\alpha}|VTV^{-1}T^{\dagger}V|R_{\beta}\rangle&=&\sum_i\langle R_{\alpha}|VT|R_i\rangle\langle R_i|VV^{-1}T^{\dagger}V|R_{\beta}\rangle=\sum_i\langle R_{\alpha}|VT|R_i\rangle\langle R_i|T^{\dagger}V|R_{\beta}\rangle
\nonumber\\
&=&\sum_i\langle R_{\alpha}|VT|R_i\rangle\langle R_{\beta}|VT|R_i\rangle^*.
\label{M51}
\end{eqnarray}
The pseudo-unitarity condition thus takes the form
\begin{eqnarray}
&&\langle R_{\alpha}|VT|R_{\beta}\rangle-\langle R_{\beta}|VT|R_{\alpha}\rangle^*=-i\sum_i\langle R_{\alpha}|VT|R_i\rangle\langle R_{\beta}|VT|R_i\rangle^*,
\nonumber\\
&&\langle L_{\alpha}|T|R_{\beta}\rangle-\langle L_{\beta}|T|R_{\alpha}\rangle^*=-i\sum_i\langle L_{\alpha}|T|R_i\rangle\langle L_{\beta}|T|R_i\rangle^*.
\label{M52}
\end{eqnarray}
We recognize this pseudo-unitarity condition as having none other than the same generic form that appears in the standard unitarity situation, save only that we have replaced the Hermitian conjugates of right-eigenvectors of the Hamiltonian by left-eigenvectors. 

In (\ref{M52}) all terms are to be calculated at energy $E+i\epsilon$. From the Schwarz reflection principle (which holds above threshold since $T(E)$ is real below threshold) we can set 
\begin{eqnarray}
\langle R_{\beta}|VT(E+i\epsilon)|R_{\alpha}\rangle^*=\langle R_{\beta}|VT(E-i\epsilon)|R_{\alpha}\rangle.
\label{M53}
\end{eqnarray}
We can thus rewrite (\ref{M52}) as
\begin{eqnarray}
&&\langle R_{\alpha}|VT(E+i\epsilon)|R_{\beta}\rangle-\langle R_{\beta}|VT(E-i\epsilon)|R_{\alpha}\rangle=-i\sum_i\langle R_{\alpha}|VT(E+i\epsilon)|R_i\rangle\langle R_{\beta}|VT(E+i\epsilon)|R_i\rangle^*,
\nonumber\\
&&\langle L_{\alpha}|T(E+i\epsilon)|R_{\beta}\rangle-\langle L_{\beta}|T(E-i\epsilon)|R_{\alpha}\rangle=-i\sum_i\langle L_{\alpha}|T(E+i\epsilon)|R_i\rangle\langle L_{\beta}|T(E+i\epsilon)|R_i\rangle^*.
\label{K54}
\end{eqnarray}
With the right-hand side of (\ref{K54}) being given by $-i$ times a positive definite quantity when $\alpha=\beta$, we see that when a Hamiltonian is similarity equivalent to a Hermitian theory, cut discontinuities have the same signs as  they do in standard unitary theories, with probability conservation above scattering thresholds thus being implemented in the standard way. When a theory is not similarity equivalent to a Hermitian theory there is no automatic positivity requirement and cut discontinuities are not immediately obliged to be positive.

As derived, (\ref{K54}) is an exact, all-order relation, and all that goes into its derivation is that the Hamiltonian obey $VHV^{-1}=H^{\dagger}$, that the time evolution operator is $U=e^{-iHt}$, and that $H$ is similarity equivalent to a Hermitian Hamiltonian. For the $-\lambda\bar{\phi}^4$ theory of interest to us here the lowest order term in $T$ in a two-particle to two-particle scattering amplitude is the tree approximation four-point vertex graph with strength $\lambda$ (shaped like the letter ${\rm X}$). Since the right-hand side of (\ref{K54}) thus must begin in order $\lambda^2$ the conventional tree approximation graph has no discontinuity. (As we will see below, through taking matrix elements in the appropriate $V$-operator  based states the tree approximation graph will in fact acquire an unconventional imaginary part, one that will prove crucial below.) With the right-hand side of (\ref{K54})  beginning in order $\lambda^2$ in the $-\lambda \bar{\phi}^4$ theory, the lowest order graph that could have a discontinuity would have to be of order $\lambda^2$, viz. precisely the one loop graph $\Pi(P^{\mu}P_{\mu})$ discussed above. To get the overall phases we recall that the time evolution operator can be written as the time-ordered product
\begin{eqnarray}
U(f,i)=I+\sum_{n=1}^{\infty}\frac{(-i)^n}{n!}\int _i^fd^4x_1...d^4x_nT\left[H_I(x_1)...H_I(x_n)\right],
\label{K55}
\end{eqnarray}
where $H_I$ is the interaction Hamiltonian density. With $H_I$ being of order $\lambda$ we can symbolically set $S=U(t_f=\infty,t_i=-\infty)=1-i\lambda \alpha-\lambda^2\beta$, and thus $T=\lambda \alpha-i\lambda^2\beta$. With $\lambda$ being taken to be real,  when $V$ is Hermitian we thus obtain $T-T^*=-i\lambda^2[\beta(E+i\epsilon)-\beta(E-i\epsilon)]=-iTT^*=-i\lambda^2\alpha(E+i\epsilon)\alpha(E+i\epsilon)^*$, with pseudo-unitarity requiring that the coefficient $\beta(E+i\epsilon)-\beta(E-i\epsilon)$ be positive. In addition we note that since the right-hand side of (\ref{K54}) is quadratic in $T$ it contains two (suppressed) four-momentum conserving delta functions. However, the left-hand side of (\ref{K54}) is linear in $T$ and thus seemingly only contains one. As we show below, a second delta function is generated by the loop diagram integration itself. We thus now proceed to an evaluation of the loop diagram.

\subsection{Loop cut discontinuities}

To evaluate the one loop $\Pi(P^{\mu}P_{\mu})$ of (\ref{N30}) in the timelike region we set $P_{\mu}=(p_0,0,0,0)$ where $p_0$ is taken to be positive, with the Feynman causal propagator then taking the form

\begin{eqnarray}
\!\Pi(p_0)=\frac{i\lambda^2}{(M_1^2-M_2^2)^2} \int \frac{d^4k}{(2\pi)^4}\left[\frac{1}{k_0^2-E_2^2+i\epsilon}-\frac{1}{k_0^2-E_1^2+i\epsilon}\right]\left[\frac{1}{(k_0+p_0)^2-E_2^2+i\epsilon}-\frac{1}{(k_0+p_0)^2-E_1^2+i\epsilon}\right],
\label{K56}
\end{eqnarray}
where $E_i^2=\bar{k}^2+M_i^2$. We write $\Pi(p_0)$ symbolically as 
 \begin{eqnarray}
 \Pi(p_0)=\frac{\lambda^2}{(M_1^2-M_2^2)^2}\left[\Pi(2,2)-\Pi(1,2)-\Pi(2,1)+\Pi(1,1)\right],
\label{K57}
 \end{eqnarray}
with $\Pi(2,2)+\Pi(1,1)$ being standard contributions of the type that one obtains in Hermitian field theories and $-\Pi(1,2)-\Pi(2,1)$ being non-standard contributions that one obtains because of the relative minus sign in $1/(k^2-M_2^2)-1/(k^2-M_1^2)$.

We evaluate the $k_0$ integration using the Feynman contour, and on closing the contour below the real $k_0$ axis in a clockwise direction obtain for the typical $\Pi(1,2)$ component
 \begin{eqnarray}
 &&\Pi(1,2)=i \int \frac{d^4k}{(2\pi)^4}\frac{1}{k_0^2-E_1^2+i\epsilon}\frac{1}{(k_0+p_0)^2-E_2^2+i\epsilon}
 \nonumber\\
&& =\int \frac{d^3k}{(2\pi)^3}\left(\frac{1}{2E_1[(p_0+E_1)^2-E_2^2+i\epsilon)]}+\frac{1}{2E_2[(p_0-E_2)^2-E_1^2+i\epsilon]}\right)
 \nonumber\\
 &&=\int_0^{\infty}  \frac{dk k^2}{2\pi^2} \frac{1}{4E_1E_2}\left(\frac{1}{p_0+E_1-E_2+i\epsilon}-\frac{1}{p_0+E_1+E_2+i\epsilon}
 +\frac{1}{p_0-E_2-E_1+i\epsilon}-\frac{1}{p_0-E_2+E_1+i\epsilon}\right)
 \nonumber\\
 &&=\int_0^{\infty} \frac{dk k^2}{2\pi^2}\frac{1}{4E_1E_2}\left(-\frac{1}{p_0+E_1+E_2+i\epsilon}
 +\frac{1}{p_0-E_2-E_1+i\epsilon}\right)
 \nonumber\\
 &&=\int_0^{\infty} \frac{dk k^2}{2\pi^2}\frac{1}{4E_1E_2}\bigg{(}-PP\left[\frac{1}{p_0+E_1+E_2}\right]+i\pi\delta(p_0+E_1+E_2)
 \nonumber\\
 &&+PP\left[\frac{1}{p_0-E_2-E_1}\right]-i\pi\delta(p_0-E_1-E_2)\bigg{)},
\label{K58}
 \end{eqnarray}
where $PP$ denotes the Cauchy principal part, and where we have kept the $i\epsilon$ in order to determine the structure of $\Pi(1,2)$ in the complex $p_0$ plane. 

Under a $CPT$ transformation the $\Pi(1,2)$ amplitude would be complex conjugated and $p_0$ would be replaced by $-p_0$. On noting that $\Pi(1,2,p_0)=\Pi(1,2,-p_0)^*$, we see that the theory is $CPT$ symmetric and that the condition $UCPTU[CPT]^{-1}=I$ is obeyed identically. With $\Pi(1,2,p_0)=\Pi(1,2,-p_0)^*$ we see that even in a non-Hermitian theory, and even with a $1/(k^2-M_2^2)-1/(k^2-M_1^2)$ propagator, we still have standard $CPT$ microreversibility.

With $p_0$ being taken to be positive, of the two delta functions that appear in (\ref{K58}) only $\delta(p_0-E_1-E_2)$ contributes, while at the same time the integration over the $PP[1/(p_0+E_1+E_2)]$ term becomes an ordinary integral, with (\ref{K58}) reducing to
 \begin{eqnarray}
 \Pi(1,2)=\int_0^{\infty} \frac{dk k^2}{8\pi^2}\frac{1}{E_1E_2}\bigg{(}-\frac{1}{p_0+E_1+E_2} +PP\left[\frac{1}{p_0-E_2-E_1}\right]-i\pi\delta(p_0-E_1-E_2)\bigg{)}. 
\label{K59}
 \end{eqnarray}
For the delta function term we see that $p_0$ is constrained to obey  $p_0=(\bar{k}^2+M_1^2)^{1/2}+(\bar{k}^2+M_2^2)^{1/2}$. (This is just the energy conservation condition for an incoming pair of particles with total four-momentum $(p_0,0,0,0)$ and two outgoing particles with four-momenta $((\bar{k}^2+M_1^2)^{1/2},\bar{k})$ and $((\bar{k}^2+M_2^2)^{1/2},-\bar{k})$, and is, as noted above, needed to balance delta functions in the non-linear $T-V^{-1}T^{\dagger}V=-iTV^{-1}T^{\dagger}V$.) Since the delta function condition can be satisfied for all allowed $k=|\bar{k}|$, we thus obtain a cut in the $p_0$ plane beginning at $p_0=M_1+M_2$. Similarly, $\Pi(1,1)$ and $\Pi(2,2)$ have branch points at $p_0=2M_1$, $p_0=2M_2$. This is just as should be the case if we cut the intermediate lines in the loop diagram on shell. Thus,  as we see, even in the non-Hermitian case, the cutting rules give the locations of branch points and branch cuts.
 
 Solving for the value of $k$ that satisfies $p_0=E_1+E_2$ we obtain
 \begin{eqnarray}
 (p_0^2-2k^2-M_1^2-M_2^2)^2=4(k^2+M_1^2)(k^2+M_2^2), 
\label{K60}
 \end{eqnarray}
from which it follows that
 \begin{eqnarray}
 4p_0^2k^2=p_0^4-2p_0^2(M_1^2+M_2^2)+M_1^4+M_2^4-2M_1^2M_2^2,
\label{K61}
 \end{eqnarray}
 with solution $k=\alpha$ where 
 \begin{eqnarray}
 2p_0\alpha=[p_0^2-(M_1+M_2)^2]^{1/2}[p_0^2-(M_1-M_2)^2]^{1/2}. 
\label{K62}
 \end{eqnarray}
On recalling that $\int dk g(k)\delta(f(k))= g(\alpha)/f^{\prime}(\alpha)$ where $f(\alpha)=0$, evaluating the delta function term in (\ref{K59}) gives 
 \begin{eqnarray}
 \Pi(1,2,\delta)=-\frac{i}{8\pi}\frac{1}{(\alpha^2+M_1^2)^{1/2}(\alpha^2+M_2^2)^{1/2}}\bigg{(}\frac{(\alpha^2+M_1^2)^{1/2}(\alpha^2+M_2^2)^{1/2}}{\alpha[(\alpha^2+M_1^2)^{1/2}+(\alpha^2+M_2^2)^{1/2}]}\bigg{)}=-\frac{i}{8\pi}\frac{\alpha}{p_0}.
\label{K63}
 \end{eqnarray}
The complete delta function contribution to $\Pi(p_0)$ from $\Pi(2,2)-\Pi(1,2)-\Pi(2,1)+\Pi(1,1)$ is thus given by
 \begin{eqnarray}
 \Pi(p_0,\delta)&=&-\frac{i\lambda^2}{8\pi(M_1^2-M_2^2)^2}\int dkk^2\left(\frac{1}{E_2^2}\delta(p_0-2E_2)+\frac{1}{E_1^2}\delta(p_0-2E_1)-\frac{2}{E_1E_2}\delta(p_0-E_1-E_2)\right)
 \nonumber\\
 &=&-\frac{i\lambda^2}{8\pi(M_1^2-M_2^2)^2}\bigg{(}\theta(p_0-2M_2)\frac{(p_0^2-4M_2^2)^{1/2}}{2p_0}+\theta(p_0-2M_1)\frac{(p_0^2-4M_1^2)^{1/2}}{2p_0}
 \nonumber\\
&-&\theta(p_0-M_1-M_2)\frac{[p_0^2-(M_1+M_2)^2]^{1/2}[p_0^2-(M_1-M_2)^2]^{1/2}}{p_0^2}\bigg{)},~
\label{K64}
 \end{eqnarray}
to nicely exhibit square root  branch points at the required places. (For comparison we note that for the standard $1/(k^2-M_2^2)+1/(k^2-M_1^2)$ propagator the coefficients of all of the theta functions would be same as that of the $\theta(p_0-2M_1)$ term.)

To make contact with the $T$ matrix we recall that in a Hermitian theory with a propagator of the form $1/(E-H)$, the $T$ matrix behaves as $1/(E-H+i\epsilon)$. Thus, in analog to (\ref{K54}), in a Hermitian theory the T matrix discontinuity is given by
 \begin{eqnarray}
 T(E+i\epsilon)-T(E-i\epsilon)=-2\pi i\delta(E-H),
\label{K65}
 \end{eqnarray}
while from the Hermitian theory $SS^{\dagger}=I$ relation we would obtain
 \begin{eqnarray}
 T(E+i\epsilon)-T(E-i\epsilon)= -iT(E+i\epsilon)T^*(E+i\epsilon).
\label{K66}
 \end{eqnarray}
Now as introduced, in the non-Hermitian fourth-order derivative case we can identify $T=i(S-I)$ with $\Pi(p_0)$. Comparing with (\ref{K64}) we see that the $\Pi(2,2)+\Pi(1,1)$ contribution has the positive standard signature that would occur in a Hermitian theory, while the $-\Pi(1,2)-\Pi(2,1)$ contribution has a non-standard negative signature. Now we had constructed the $V$ operator of the PU theory to be the Hermitian $e^{-Q}$, and thus (\ref{M52}) and (\ref{K54}) should be obeyed and all cut discontinuities should have positive signature. And yet the one associated with the $-\Pi(1,2)-\Pi(2,1)$ contribution does not. We now clarify the point and show that despite this pseudo-unitarity is in fact obeyed.

\subsection{Reconciling negative cut discontinuities with pseudo-unitarity}

In adding on a $-\lambda y^4$ term or a $-\lambda\bar{\phi}^4$ term to the Lagrangian we are replacing a Hamiltonian such as $\bar{H}$ of (\ref{G15}) by $\bar{H}+\lambda y^4$. Now since we have added on a Hermitian $\lambda y^4$, we might expect that  this should not affect Hermiticity considerations, but in fact it does since $\bar{H}$ is not Hermitian. Specifically, we introduced the Hermitian $Q$ operator of (\ref{G26}) that effects $e^{-Q}\bar{H}e^{Q}=\bar{H}^{\dagger}$ while generating an  $H^{\prime}=e^{-Q/2}\bar{H}e^{+Q/2}$ that is Hermitian. However under this latter transformation $y$ transforms into $y^{\prime}=e^{-Q/2}ye^{Q/2}=y\cosh\theta+i(\alpha/\beta)^{1/2}p\sinh\theta$ where $\theta=(\alpha\beta)^{1/2}/2$. With $y$ and $p$ both being Hermitian, we see that $y^{\prime}$ is not Hermitian. Thus after transforming with $Q$ we find that $e^{-Q/2}[\bar{H}+\lambda y^4]e^{Q/2}=\bar{H}^{\prime}+\lambda y^{\prime 4}$ is not Hermitian, and that the Hermiticity of $\lambda y^4$ has effectively been transformed away. In passing we note that if $y$ were not to transform,  the Hamiltonian would then be of the standard anharmonic oscillator form $\bar{H}^{\prime}+\lambda y^4=p^2/2+q^2/2\omega_1^2+\omega_1^2x^2/2+\omega_1^2\omega_2^2y^2/2+\lambda y^4$ with propagator $1/(E^2-\omega_1^2)+1/(E^2-\omega_2^2)$, an entirely different theory, one not similarity equivalent to $\bar{H}+\lambda y^4$ at all.

Now the derivation of (\ref{M52}) and (\ref{K54}) required that the requisite $V$ be of the form  $B^{\dagger}B$ where the $BHB^{-1}$ transformation would bring the Hamiltonian to a Hermitian form. However, this is not the case for $\bar{H}+\lambda y^4$ even though the free theory $e^{-Q/2}$ is Hermitian. Consequently, the pseudo-unitarity condition given in (\ref{M52}) and (\ref{K54}) does not apply and cut discontinuities are not obliged to be positive. The calculation of the loop diagram that we have presented is thus not in conflict with either standard unitarity or the pseudo-unitarity condition (\ref{M52}) and (\ref{K54}) as neither applies.

However, that does not therefore mean that there is a negative cut discontinuity in the theory, since in the presence of the $\lambda y^4$ term the operator $e^{-Q}$ is no longer the relevant $V$ operator. Rather, we need an operator that brings the entire $\bar{H}+\lambda y^4$ to a Hermitian form, rather than just $\bar{H}$ itself. Perturbatively, we would only need to do this to lowest order in $\lambda$. So let us introduce a new operator $\hat{B}=(1+\lambda A)e^{-Q/2}$, with $\hat{B}^{-1}=e^{Q/2}(1-\lambda A)$, and $\hat{B}^{\dagger}=e^{-Q/2}(1+\lambda A^{\dagger})$ to lowest order in $\lambda$. To lowest order in $\lambda$ let us identify 
\begin{eqnarray}
\hat{V}=\hat{B}^{\dagger}\hat{B}=e^{-Q/2}(1+\lambda A+\lambda A^{\dagger})e^{-Q/2},\qquad
\hat{V}^{-1}=e^{Q/2}(1-\lambda A-\lambda A^{\dagger})e^{Q/2}.
\label{M67}
\end{eqnarray}
If $\hat{B}(\bar{H}+\lambda y^4)\hat{B}^{-1}$ is Hermitian, it is equal to $\hat{B}^{\dagger -1}(\bar{H}^{\dagger}+\lambda y^4)\hat{B}^{\dagger}$, with  $\hat{V}=\hat{B}^{\dagger}\hat{B}$ thus implementing $\hat{V}(\bar{H}+\lambda y^4)\hat{V}^{-1}=\bar{H}^{\dagger}+\lambda y^4$. Similarly, if $\hat{V}$ implements $\hat{V}(\bar{H}+\lambda y^4)\hat{V}^{-1}=\bar{H}^{\dagger}+\lambda y^4$, then $\hat{B}(\bar{H}+\lambda y^4)\hat{B}^{-1}$ is Hermitian. Thus to fix $A$ we just need to satisfy
\begin{eqnarray}
&&\hat{V}(\bar{H}+\lambda y^4)\hat{V}^{-1}=e^{-Q}\bar{H}e^{Q}+ 
\lambda e^{-Q/2}(A+A^{\dagger})e^{-Q/2}\bar{H}e^{Q}
-\lambda e^{-Q}\bar{H}e^{Q/2}(A+A^{\dagger})e^{Q/2}
+\lambda e^{-Q} y^4e^{Q}
\nonumber\\
&&
=\bar{H}^{\dagger}+\lambda y^4=e^{-Q}\bar{H}e^{Q}+\lambda y^4,
\label{M68}
\end{eqnarray}
with $A$ being the solution to
\begin{eqnarray}
e^{Q/2}(A+A^{\dagger})e^{-Q/2}\bar{H}-\bar{H}e^{Q/2}(A+A^{\dagger})e^{-Q/2}
=e^{Q}y^4e^{-Q}-y^4.
\label{M69}
\end{eqnarray}

While it does not appear to be possible to solve for $A$ analytically, we note that in lowest order perturbation theory the shift to any energy eigenvalue of the free $\bar{H}$ theory is given by $\lambda \langle n_L|y^4|n_R\rangle=\lambda \langle n_R|e^{-Q}y^4|n_R\rangle$. With such a shift being real (since it must contain an even number of the $i$ factors given in (\ref{G16}) for $y(t)$), to lowest order in $\lambda$ we see that the eigenvalues of $\bar{H}+\lambda y^4$ are all real. With the eigenvectors being complete (something that cannot change in perturbation theory) $\bar{H}+\lambda y^4$ must be similarity equivalent to a Hermitian Hamiltonian. Therefore there must be a $\hat{B}$ and a $\hat{V}$, and there therefore must be a solution for $A$. Therefore according to (\ref{K54}) there  thus must be a pseudo-unitarity relation for $\hat{V}$ of the form 
\begin{eqnarray}
\langle R_{\alpha}|\hat{V}\hat{T}(E+i\epsilon)|R_{\beta}\rangle-\langle R_{\beta}|\hat{V}\hat{T}(E-i\epsilon)|R_{\alpha}\rangle=-i\sum_i\langle R_{\alpha}|\hat{V}\hat{T}(E+i\epsilon)|R_i\rangle\langle R_{\beta}|\hat{V}\hat{T}(E+i\epsilon)|R_i\rangle^*,
\label{M70}
\end{eqnarray}
where $\hat{T}$ is the $T$ matrix for $\bar{H}+\lambda y^4$. 

Even though  (\ref{M70}) guarantees that there be no negative cut discontinuities in $\hat{T}$, it is instructive to see how the negative cut discontinuity in the loop diagram is actually  cancelled since it is not immediately apparent. With the negative cut discontinuity in the loop diagram being of order $\lambda^2$, we need to generate an additional term of order $\lambda^2$ in (\ref{M70}), and not only that, it would need to be imaginary. With the difference between $\hat{V}$ and $e^{-Q}$ being of order $\lambda$, on the left-hand side of (\ref{M70}) we would need a contribution from $\hat{T}$ that is also of order $\lambda$. Since in general one can write the $T$ matrix for a Hamiltonian $H_0+H_I$ as $T=H_I+H_1(E-H_0+i\epsilon)^{-1}H_I+....$, the term of relevance for us in $\hat{T}$ would be  $\hat{T}=\lambda y^4$. As per the structure of (\ref{M52}), the relevant term of order $\lambda^2$ in the left-hand side of (\ref{M70}) is thus given by
\begin{eqnarray}
&&\langle R_{\alpha}|\hat{V}\hat{T}|R_{\beta}\rangle-\langle R_{\beta}|\hat{V}\hat{T}|R_{\alpha}\rangle^*
\nonumber\\
&&=\lambda^2\sum_i\left[\langle R_{\alpha}|e^{-Q/2}(A+A^{\dagger})e^{-Q/2}|R_i\rangle\langle R_i|e^{-Q}y^4|R_{\beta}\rangle-
\left(\langle R_{\alpha}|e^{-Q/2}(A+A^{\dagger})e^{-Q/2}|R_i\rangle\langle R_i|e^{-Q}y^4|R_{\beta}\rangle\right)^*\right]
\nonumber\\
&&=\lambda^2\sum_i\left[\langle R_{\alpha}|e^{-Q/2}(A+A^{\dagger})e^{-Q/2}|R_i\rangle-
\langle R_{\alpha}|e^{-Q/2}(A+A^{\dagger})e^{-Q/2}|R_i\rangle^*\right]\langle R_i|e^{-Q}y^4|R_{\beta}\rangle
\nonumber\\
&&=\lambda^2\sum_i\left[\langle L_{\alpha}|e^{Q/2}(A+A^{\dagger})e^{-Q/2}|R_i\rangle-
\langle L_{\alpha}|e^{Q/2}(A+A^{\dagger})e^{-Q/2}|R_i\rangle^*\right]\langle L_i|y^4|R_{\beta}\rangle,
\label{M71}
\end{eqnarray}
a term which nicely has an imaginary part of order $\lambda^2$. We note that to order $\lambda^2$ all the states that appear in  (\ref{M71}) have to be $\lambda$ independent. These states thus have to be eigenstates of $\bar{H}$, and as noted above, for them matrix elements of $y^4$ are real. Consequently, in (\ref{M71}) we were able to set 
$\langle R_i|e^{-Q}y^4|R_{\beta}\rangle=\langle R_i|e^{-Q}y^4|R_{\beta}\rangle^*$. Also we note that while $y^4$ could connect an external two-particle state to an intermediate state with $0$, $2$, $4$ or $6$ particles,  the only loop diagram negative cut discontinuity that would need to be cancelled in order $\lambda^2$ is the one associated with the two-particle to two-particle $\Pi(1,2)$. (A two-particle to four-particle loop for instance would be of order $\lambda^3$.) Hence in (\ref{M71}) all states of interest to us here are two-particle states.

With the insertion of a summation of a complete set of on-shell intermediate two-particle states that we have made in (\ref{M71}), (\ref{M71}) has exactly the same $\delta(p_0-2E_2)$,  $\delta(p_0-2E_1)$,  $\delta(p_0-E_1-E_2)$ set of delta functions as the one loop $\Pi(p_0,\delta)$ given in (\ref{K64}). When the delta function terms in (\ref{K64}) and (\ref{M71}) are combined on the left-hand side of (\ref{M70}) ((\ref{M70}) contains all $\lambda^2$ contributions), the coefficients of each of  the three of them are then of the standard form of $-i$ times a positive quantity. Thus the introduction of $\hat{V}$ does not affect the locations of the branch points or the presence of branch cuts, but does affect the magnitudes and signs of the discontinuities across the cuts.

As we see, the essence of the effect is that  while both $\langle R|\lambda y^4|R\rangle$ and $\langle R|e^{-Q}\lambda y^4|R\rangle$ would be real and first order in $\lambda$, $\langle R|\hat{V}\lambda y^4|R\rangle$ contains a term that is both second order in $\lambda$ and complex. The effect is entirely due to the states in which we evaluate matrix elements and not due to the vertex itself, with the vertex still being the bare point-coupled $\lambda y^4$. For the vertex itself the contribution of interest to us is an order $\lambda$ tree approximation contribution, with a second power of $\lambda$ being generated through the fact that the Hamiltonian is  not Hermitian, with left-eigenvectors  not being the Hermitian conjugates of right-eigenvectors.

In addition, not only does the generation of this second power of $\lambda$ lead to the cancellation of the negative cut discontinuity, it involves no momentum dependence other than the energy conserving delta functions, and thus does not affect the asymptotic spacelike behavior of the theory. Thus even while we cancel the negative cut discontinuity, we lose none of the good ultraviolet behavior of the $1/(k^2-M_2^2)-1/(k^2-M_1^2)$ propagator. Moreover, this same cancellation generalizes to higher order graphs with negative-signatured discontinuities in loop graphs of order $\lambda^{2n}$ being cancelled by a tree graph with a $\hat{V}$ contribution of order $\lambda^{2n-1}$ and a four-point vertex contribution of order $\lambda$. Thus with the inclusion of the $\lambda$ dependent term in $\hat{V}$ all cut discontinuities in Feynman diagrams will obey the positivity that is required by the pseudo-unitarity  condition given in (\ref{M70}). It is thus through the very unusual behavior of the tree approximation graph contribution then that unitarity is secured, with there being no observable negative probabilities.

To conclude, we see that having a $1/(k^2-M_2^2)-1/(k^2-M_1^2)$ propagator does not lead to nonconservation of probability. We also note that the very mechanism that leads to the partial fraction decomposition of the $1/[(k^2-M_2^2)(k^2-M_1^2)]$ propagator in the first place, namely the existence of a fourth-order derivative theory based on $I_{S}$, also leads to the theory being non-Hermitian but $PT$ symmetric. And this in turn is then responsible for maintaining the unitarity in  loop diagrams that provides for the resolution of the very problem that this propagator appears to create. Thus by possessing this wisdom, the fourth-order theory is able to take care of itself. As we shall see below, the wisdom associated with the  fourth-order theory will itself need to be modified in the equal-frequency  or equal mass limits, since the partial fraction decompositions in (\ref{G23}) and (\ref{M82}) below become undefined when $\omega_1=\omega_2$ and $M_1=M_2$.

\section{The Pauli-Villars regulator scheme and the Lee-Wick mechanism}

In trying to regulate the asymptotic momentum behavior of  Feynman diagrams, Pauli and Villars \cite{Pauli1949} suggested that one replace $D(k)=1/(k^2-M_2^2)$ by $D(k)=1/(k^2-M_2^2)-1/(k^2-M_1^2)$. As conceived by Pauli and Villars it was necessary that both of the $1/(k^2-M_2^2)$ and $1/(k^2-M_1^2)$ propagators couple to vertices with the same relative sign. The two propagators would act as mirror images of each other, to thus be associated with two independent and decoupled second-order derivative actions
\begin{eqnarray}
I_{S_2}+I_{S_1}=\int d^4x[\frac{1}{2}\partial_{\mu}\phi_2\partial^{\mu}\phi_2-\frac{1}{2}M_2^2\phi_2^2-\lambda\phi_2^4]+\int d^4x[\frac{1}{2}\partial_{\mu}\phi_1\partial^{\mu}\phi_1-\frac{1}{2}M_1^2\phi_1^2-\lambda\phi_1^4].
\label{M72}
\end{eqnarray}
However, in order to generate the relative minus sign in $D(k)=1/(k^2-M_2^2)-1/(k^2-M_1^2)$ so as to effect the regulation cancellation, one would have to take $\phi_2$ to be quantized with positive norm and $\phi_1$ to be quantized in a  negative norm Krein space. Thus in this case the relative minus sign in $D(k)$ really would be due to using an indefinite metric.

The objective of Pais and Uhlenbeck was to see whether one could generate the same $D(k)=1/(k^2-M_2^2)-1/(k^2-M_1^2)$ propagator from an action involving a single field, namely the fourth-order derivative $I_{S}$ associated with a single field $\phi$. However, it was thought that this $I_S$ action would itself lead to either negative energy states or negative norm states. And it was only with the work of \cite{Bender2008a,Bender2008b} that  it was shown that there were in fact no states of negative norm or negative energy in the PU two-oscillator theory after all. Left open was the question of how could one avoid negative probabilities once one switched on a $-\lambda\phi^4$ interaction. Since we have now addressed this question in this paper, it appears to us that one does not need to consider the action based on (\ref{M72}) at all,   and that via a fourth-order derivative $I_S$-based theory one can have Pauli-Villars regulation without negative norm states or loss of unitarity.

With there being no unitarity problem for both the free and interacting fourth-order derivative theories, there is no need to invoke the Lee-Wick mechanism \cite{Lee1969} for the  $D(k)=1/(k^2-M_2^2)-1/(k^2-M_1^2)$ propagator. In this mechanism Lee and Wick took the $D(k)$ propagator to be associated with a non-unitary, indefinite metric theory of the form exhibited in (\ref{M72}), and then found that they could obtain unitarity if they took the masses to be complexified into a complex conjugate pair according to $\tilde{D}(k)=1/(k^2-M^2+iN^2)-1/(k^2-M^2-iN^2)$. Now we had noted above that for any $H$ with antilinear symmetry, $H$ and $H^{\dagger}$ are related by $VHV^{-1}=H^{\dagger}$. With $H$ and $H^{\dagger}$ thus having the same set of eigenvalues, the eigenvalues of $H$ can either be real or appear in complex conjugate pairs. Thus not only does $D(k)=1/(k^2-M_2^2)-1/(k^2-M_1^2)$ fall into the class of antilinearly symmetric theories, so does  $\tilde{D}(k)=1/(k^2-M^2+iN^2)-1/(k^2-M^2-iN^2)$ \cite{Mannheim2015b}. Thus both of the $D(k)$ and $\tilde{D}(k)$ propagators are associated with unitary theories, both of the propagators are associated with non-Hermitian Hamiltonians that possess an antilinear symmetry,  and there is no need to complexify $D(k)$ in order to obtain unitarity as not only does $D(k)$ already possess it, the theory continues to possess it in the presence of interactions. Also in regard to the Lee-Wick mechanism, we recall that in the literature there have been some concerns expressed as to whether or not it is causal. To this end we note that through use of the fourth-order propagator equation given in (\ref{M82}) below both of the $D(k)$ and $\tilde{D}(k)$ propagators have been shown to be  causal \cite{Mannheim2015b}. (Both $1/(k^2-M_2^2)$ and $1/(k^2-M_1^2)$ are separately causal if one uses the Feynman contour for each one, and the pole structure does not change if one uses $1/(k^2-M_2^2)$ and $-1/(k^2-M_1^2)$.) Moreover, in the presence of interactions, if the theory remains similarity equivalent to a Hermitian theory causality could not be lost.  (If interactions were to move poles from the lower-right quadrant in the complex $k_0$ plane to the upper-right quadrant, the Feynman contour would have to be deformed so that these upper-right-quadrant poles would be enclosed in the contour integration.) Thus when the Lee-Wick mechanism is associated with a fourth-order derivative theory its  causality can be secured.

\section{The equal-frequency limit of the PU theory}

While we can use the $Q$ operator to bring $\bar{H}$ to the Hermitian $\bar{H}^{\prime}$ given in (\ref{N28}), we note that the factor $\alpha$ in (\ref{G26}) becomes undefined in the equal-frequency limit in which $\omega_1$ and $\omega_2$ become equal. In addition, in this limit the commutation relations in (\ref{G17}) also become singular. Moreover, if we make the partial fraction decomposition of the propagator given in (\ref{G23}), we see that in the limit $\omega_1^2\rightarrow \omega_2^2$, the decomposition becomes undefined, with the limit being singular. The decomposition of the unequal-frequency propagator into two separate pole terms thus does not provide a reliable guide as to the structure of the equal-frequency theory as the limit is singular. In fact, the equal-frequency theory is qualitatively different from the unequal-frequency theory, with it being found \cite{Bender2008b} that in the equal-frequency limit $\bar{H}$ cannot be brought to a Hermitian form at all, as it instead becomes of non-diagonalizable Jordan-block form. 

To see explicitly what happens in this limit,  we set $\omega_1=\omega+\epsilon$, $\omega_2=\omega-\epsilon$, and rewrite $\bar{G}(t)$ of (\ref{G24}) as
\begin{eqnarray}
\bar{G}(t)=-\frac{i}{8 \omega\epsilon}\left[-\frac{e^{-i(\omega+\epsilon)t}}{\omega+\epsilon}
+\frac{e^{-i(\omega-\epsilon)t}}{\omega-\epsilon}\right].
\label{M73}
\end{eqnarray}
On taking the $\epsilon\rightarrow 0$ limit we obtain
\begin{eqnarray}
\bar{G}(t)=-\frac{ie^{-i\omega t}}{8 \omega\epsilon}\left[-\frac{(1-i\epsilon t)}{\omega+\epsilon}
+\frac{(1+i\epsilon t)}{\omega-\epsilon}\right].
\label{M74}
\end{eqnarray}
Thus we obtain
\begin{eqnarray}
\bar{G}(t)=-\frac{ie^{-i\omega t}}{4 \omega^3}[1+i\omega t],
\label{M75}
\end{eqnarray}
and note that it was only because the $N_1$ and $N_2$ terms in (\ref{G25}) did have opposite signs that we were able to cancel the overall $1/\epsilon$ factor in (\ref{M74}) and get a finite limiting value in (\ref{M75}). Inspection of (\ref{M75}) shows it to consist of one standard, stationary,  oscillating $e^{-i\omega t}$ term together with a non-standard, non-stationary $te^{-i\omega t}$ term that grows in time. Now a Jordan-block Hamiltonian cannot be diagonalized because it has an incomplete set of eigenstates. The two stationary eigenstates $|1\rangle$ and $|2\rangle$ that existed before we took the equal-frequency limit, reemerge as one stationary state and one non-stationary state, with the non-stationary state being a solution to the time-dependent Schr\"odinger equation but not to the time-independent one \cite{Bender2008b}.

On Fourier transforming $\theta(t)\bar{G}(t)$,  we obtain
\begin{eqnarray}
\bar{G}(E)=\frac{1}{4 \omega^2}\left[\frac{1 }{\omega(E-\omega+i\epsilon)}-\frac{1}{(E-\omega+i\epsilon)^2}\right],
\label{M76}
\end{eqnarray}
with (\ref{M76}) being the $\epsilon \rightarrow 0$ limit of (\ref{G25}). We recognize the $+1/[4\omega^3(E-\omega+i\epsilon)]$ term as being a one-particle pole term, and note not only that its contribution to $\bar{G}(E)$ is positive definite, but that this positivity is obtained only because we had to continue $z$ into the complex plane and replace it by $y=-iz$. While there is a relative minus sign between the two pole terms of order one in the unequal-frequency $\bar{G}(E)$, in the equal-frequency  limit its single pole term of order one has no minus sign, with the other pole term of order one being replaced by a double pole term of order two. To get the Feynman rules for the equal-frequency PU oscillator one can start with the Feynman rules for $\bar{H}$ of (\ref{G15}) or (\ref{G19}) and take the $\epsilon \rightarrow 0$ limit. However, since non-stationary states are involved in the $\epsilon=0$ Jordan-block case, the standard cutting rules would not apply.

\section{The zero mass limit of the scalar field theory}

When we first studied the scalar action $I_S$ in (\ref{G1}) we set $\omega_1=(\bar{k}^2+M_1^2)^{1/2}$, $\omega_2=(\bar{k}^2+M_2^2)^{1/2}$, and to obtain a non-relativistic limit then dropped $|\bar{k}|$. However, there is a second limit, namely to set $M_1=0$, $M_2=0$ while retaining $|\bar{k}|$ with no non-relativistic limitation on its magnitude being needed, and then identify $\omega_1=\omega_2=|\bar{k}|=\omega$. Specifically, if we set both $M_1$ and $M_2$ equal to zero in $I_S$,  then $I_S$ and $T_{00}$ given in (\ref{G3})  take the form 
\begin{eqnarray}
I_S(M_1=M_2=0)&=&\frac{1}{2}\int d^4x\partial_{\mu}\partial_{\nu}\phi\partial^{\mu}
\partial^{\nu}\phi,
\nonumber\\
T_{00}(M_1=M_2=0)&=&-\dddot{\phi}\dot{\phi}+\dot{\phi}\nabla^2\dot{\phi}+\frac{1}{2}\ddot{\phi}^2-\frac{1}{2}\partial_i\partial_j\phi\partial^i\partial^j\phi.
\label{M77}
\end{eqnarray}
To show that $I_{S}(M_1=M_2=0)$ is analogous to the equal-frequency $I_{\rm PU}$, we initially assume this to be the case and then show that the assumption is consistent. With the unequal-frequency $H_{\rm PU}$ being given in (\ref{G11}), in the equal-frequency limit $H_{\rm PU}$ is given by
\begin{eqnarray}
&&H_{\rm PU}=\frac{p_x^2}{2}+p_zx+\omega^2x^2-\frac{1}{2}\omega^4z^2.
\label{N78}
\end{eqnarray}
Thus we can set
\begin{eqnarray}
\phi\equiv z,\qquad \dot{\phi}\equiv i[H_{\rm PU},z]=x,\qquad \ddot{\phi}\equiv i[H_{\rm PU},x]=p_x,\qquad\dddot{\phi}\equiv i[H_{\rm PU},p_x]=-p_z-2\omega^2x.
\label{M79}
\end{eqnarray}
On taking the spatial dependence of $\phi$ to be $e^{i\bar{k}\cdot \bar{x}}$ we obtain
\begin{eqnarray}
T_{00}(M_1=M_2=0)&\equiv&\frac{p_x^2}{2}+p_zx+\omega^2x^2-\frac{1}{2}\omega^4z^2.
\label{M80}
\end{eqnarray}
Since we recognize $T_{00}(M_1=M_2=0)$ as being the equal-frequency $H_{\rm PU}$, our identification of $I_S(M_1=M_2=0)$ with the equal-frequency $I_{\rm PU}$ is valid, and results that we obtained for the equal-frequency PU oscillator thus apply to $I_{S}(M_1=M_2=0)$, with $I_{S}(M_1=M_2=0)$ thus also being associated with a Jordan-block Hamiltonian. We also note that with $[x,z]=0$, $[p_x,z]=0$, $[p_z,z]=-i$, from (\ref{M79}) we obtain the pattern of equal time commutators 
\begin{eqnarray}
&& \delta(t)[\dot{\phi}(x),\phi(0)]=0, \qquad \delta(t)[\ddot{\phi}(x),\phi(0)]=0, \qquad \delta(t)[\dddot{\phi}(x),\phi(0)]=i\delta^4(x),
\label{M81}
\end{eqnarray}
which enforces $(\partial_t^2-\nabla^2)(\partial_t^2-\nabla^2)D(x)=-\delta^4(x)$ when $D(x)=i\langle \Omega|T(\phi(x)\phi(0))|\Omega\rangle$,  as required to be in line with (\ref{G4}). To summarize, we see that the unequal-frequency PU two-oscillator model corresponds to the non-relativistic limit of a second-order plus fourth-order derivative $I_S$ theory, while the equal-frequency PU oscillator model corresponds to a pure fourth-order derivative $I_{S}(M_1=M_2=0)$ theory with no mass terms.

While we have related $T_{00}(M_1=M_2=0)$ to $H_{\rm PU}$, as with $H_{\rm PU}$ we actually need to find analogs of (\ref{G14}) and (\ref{G15}). We thus introduce $\bar{\phi}=-i\phi$. We provide an analog of $\bar{H}$ in (\ref{A1}) below, and replace the  $D(x)=i\langle \Omega|T(\phi(x)\phi(0))|\Omega\rangle$ propagator by $\bar{D}(x)=-i\langle \Omega_L|T(\bar{\phi}(x)\bar{\phi}(0))|\Omega_R\rangle$.
For non-zero masses first, using the commutators provided in (\ref{A2}) below, the propagator obeys
\begin{eqnarray}
&&(\partial_t^2-\nabla^2+M_1^2)(\partial_t^2-\nabla^2+M_2^2)\bar{D}(x)=-\delta^4(x),
\nonumber\\
\bar{D}(x)&=&\int \frac{d^4k}{(2\pi)^4}e^{-ik\cdot x}\bar{D}(k)=-\int \frac{d^4k}{(2\pi)^4}\frac{e^{-ik\cdot x}}{(k^2-M_1^2)(k^2-M_2^2)}
\nonumber\\
&=&\int \frac{d^4k}{(2\pi)^4}\frac{1}{(M_1^2-M_2^2)}\left(-\frac{e^{-ik\cdot x}}{k^2-M_1^2}+\frac{e^{-ik\cdot x}}{k^2-M_2^2}\right),
\label{M82}
\end{eqnarray}
while for the zero mass $I_S(M_1=M_2=0)$ we have
\begin{eqnarray}
&&(\partial_t^2-\nabla^2)(\partial_t^2-\nabla^2)\bar{D}(x)=-\delta^4(x),
\nonumber\\
&&\bar{D}(x)=-\int \frac{d^4k}{(2\pi)^4}\frac{e^{-ik\cdot x}}{k^4}=-\int \frac{d^4k}{(2\pi)^4}\frac{e^{-ik\cdot x}}{(k_0^2-\bar{k}^2+i\epsilon)^2}
=-\int \frac{d^3kdE}{(2\pi)^4}\frac{e^{-iEt+i\bar{k}\cdot \bar{x}}}{(E^2-\omega^2+i\epsilon)^2}. 
\label{M83}
\end{eqnarray}
As we establish momentarily, the last expression given in (\ref{M83}) defines the contour needed for the complex $k_0=E$ plane integration.  Since the pure fourth-order propagator is of the $1/k^4$ form, it does not admit of a partial fraction decomposition, and thus suffers from none of the negative probability concerns that we had to resolve above for the second-order plus fourth-order propagator given in (\ref{G5}).

For the $1/k^4$ propagator we can rewrite the integrand in the last expression in (\ref{M83}) as 
\begin{eqnarray}
\bar{G}(E)&=&-\frac{1}{(E^2-\omega^2+i\epsilon)^2}= -\frac{1}{2\omega}\frac{d}{d\omega}
\left[\frac{1}{2\omega(E-\omega+i\epsilon)}-\frac{1}{2\omega(E+\omega-i\epsilon)}\right]
\nonumber\\
&=&\frac{1}{4 \omega^2}\left[\frac{1 }{\omega(E-\omega+i\epsilon)}-\frac{1}{(E-\omega+i\epsilon)^2}\right]-\frac{1}{4 \omega^2}\left[\frac{1 }{\omega(E+\omega-i\epsilon)}+\frac{1}{(E+\omega-i\epsilon)^2}\right].
\label{M84}
\end{eqnarray}
With (\ref{M84}) recovering (\ref{M76}) at the $E=+\omega$ pole, we confirm that the form for $1/k^4$ as $1/(k_0^2-\bar{k}^2+i\epsilon)^2$ is the correct form for the complex $k_0$ plane singularity structure of $1/k^4$.

In addition, we note that it actually is possible to construct the massless $1/k^4$ propagator as the non-singular limit of a propagator that is massive. Specifically, we set 
\begin{eqnarray}
\bar{D}(x)=-\int \frac{d^4k}{(2\pi)^4}\frac{e^{-ik\cdot x}}{k^4}=-\int \frac{d^4k}{(2\pi)^4}\frac{e^{-ik\cdot x}}{(k^2+i\epsilon)^2}=-\lim_{M^2\rightarrow 0}\int \frac{d^4k}{(2\pi)^4}\frac{d}{dM^2}\left(\frac{e^{-ik\cdot x}}{k^2-M^2+i\epsilon}\right).
\label{M85}
\end{eqnarray}
Since there are no problematic minus signs in (\ref{M85}) (or in fact anything untoward at all in (\ref{M85})), and since the $M^2\rightarrow 0$ limit in (\ref{M85}) is not singular, in Feynman diagrams we could replace the $1/k^4$ propagator by the $M^2$-dependent expression in (\ref{M85}). Then since this $M^2$-dependent expression would give no problems at all (it being nothing other than the derivative of a normal second-order propagator),  we confirm that the pure fourth-order derivative theory is free of negative norm states, and that its Feynman loop diagrams are perfectly viable.  For the $-\lambda\bar{\phi}^4$ theory for instance the massless theory one loop graph would take the form
\begin{eqnarray}
\Pi(P^{\mu}P_{\mu})&=&\lambda^2\lim_{M_1^2\rightarrow 0}\lim_{M_2^2\rightarrow 0}\frac{d}{dM_1^2}\frac{d}{dM_2^2}
\int d^4k\frac{1}{(k^2-M_1^2+i\epsilon)}\frac{1}{((k+p)^2-M_2^2+i\epsilon)},
\label{M86}
\end{eqnarray}
with no propagators with negative signs being encountered. Moreover, since for the general Feynman graph the $M_i^2\rightarrow 0$ limit ($i=1,2,3...$) is continuous, we can determine the cutting rules for the massless theory before we take the $M_i^2\rightarrow 0$ limit.

Finally, we note that despite its Jordan-block structure, the pure fourth-order derivative theory is not without physical interest as the conformal gravity theory (viz. a conformal invariant theory of gravity that is based on the action $I_{\rm W}=-\alpha_g\int d^4x (-g)^{1/2}C_{\lambda\mu\nu\tau}C^{\lambda\mu\nu\tau}$ where $C_{\lambda\mu\nu\tau}$ is the conformal Weyl tensor and $\alpha_g$ is a dimensionless coupling constant) that has been advanced as a candidate alternative to standard gravity is also a pure fourth-order derivative Jordan-block theory \cite{Mannheim2006}. With the conformal gravity theory propagator being of the highly convergent $1/k^4$ form in the ultraviolet the theory is renormalizable, and because the propagator has the same $1/k^4$ form in the infrared (there being no intrinsic mass scales in a conformal invariant theory) the conformal gravity theory is thus  ghost free, and  its perturbative loop expansion is fully viable. Conformal gravity is thus offered as a fully consistent and renormalizable quantum theory of gravity.

\section{acknowledgments} The author would like to thank Dr. G. 't Hooft for helpful comments.

\appendix
\setcounter{equation}{0}
\section{The relativistic case}

Given the analogy with the PU two-oscillator model, for the full relativistic case the Hamiltonian $\bar{H}$ and commutation relations associated with $\bar{\phi}=-i\phi$  are given by \cite{Bender2008b}
\begin{eqnarray}
\bar{H}&=&\int d^3k\bigg{[}2(M_1^2-M_2^2)(\bar{k}^2+M_1^2)\hat{a}_{1,\bar{k
}}a_{1,\bar{k}}+2(M_1^2-M_2^2)(\bar{k}^2+M_2^2)\hat{a}_{2,\bar{k}}a_{2,
\bar{k}}
\nonumber\\
&&+\frac{1}{2}(\bar{k}^2+M_1^2)^{1/2}+\frac{1}{2}(\bar{k}^2+M_2^2)^{1/2}\bigg{]},
\label{A1}
\end{eqnarray}
and
\begin{eqnarray}
&& \delta(t)[\dot{\bar{\phi}}(x),\bar{\phi}(0)]=0,\qquad \delta(t)[\ddot{\bar{\phi}}(x),\bar{\phi}(0)]=0,\qquad \delta(t)[\dddot{\bar{\phi}}(x),\bar{\phi}(0)]=-i\delta^4(x),
\nonumber\\
&&[a_{1,\bar{k}},\hat{a}_{1,\bar{k}^{\prime}}]=[2(M_1^2-M_2^2)(\bar{k}^2+
M_1^2)^{1/2}]^{-1}\delta^3(\bar{k}-\bar{k}^{\prime}),\nonumber\\
&&[a_{2,\bar{k}},\hat{a}_{2,\bar{k}^{\prime}}]=[2(M_1^2-M_2^2)(\bar{k}^2+
M_2^2)^{1/2}]^{-1}\delta^3(\bar{k}-\bar{k}^{\prime}),\nonumber\\
&&[a_{1,\bar{k}},a_{2,\bar{k}^{\prime}}]=0,\quad[a_{1,\bar{k}},\hat{a}_{2,
\bar{k}^{\prime}}]=0,\quad[\hat{a}_{1,\bar{k}},a_{2,\bar{k}^{\prime}}]=0,\quad
[\hat{a}_{1,\bar{k}},\hat{a}_{2,\bar{k}^{\prime}}]=0,
\label{A2}
\end{eqnarray}
where all relative signs are positive (we take $M_1^2>M_2^2$). The field expansion is given by
\begin{eqnarray}
\bar{\phi}(x)=\int \frac{d^3k}{(2\pi)^{3/2}}\left [-ia_{1,\bar{k}}e^{-ik^1\cdot x}+a_{2,\bar{k}}e^{-ik^2\cdot x}-i\hat{a}_{1,\bar{k}}e^{ik^1\cdot x}+\hat{a}_{2,\bar{k}}e^{ik^2\cdot x}\right].
\label{A3}
\end{eqnarray}
The left- and right-vacua obey
\begin{eqnarray}
a_{1,\bar{k}}|\Omega_R\rangle =0,\qquad a_{2,\bar{k}}|\Omega_R\rangle =0,\qquad \langle \Omega_L|\hat{a}_{1,\bar{k}}=0,\qquad \langle \Omega_L|\hat{a}_{2,\bar{k}}=0,\qquad \langle \Omega_L|\Omega_R\rangle =1,
\label{A4}
\end{eqnarray}
and the one-particle states with energies $(\bar{k}^2+
M_1^2)^{1/2}$ and $(\bar{k}^2+
M_2^2)^{1/2}$ above the ground state are given by
\begin{eqnarray}
&&|k_{R}^{i}\rangle=[2(M_1^2-M_2^2)(\bar{k}^2+
M_i^2)^{1/2}]^{1/2}\hat{a}_{i,\bar{k}}|\Omega_R\rangle,
\nonumber \\
\ &&\langle k_{L}^{i}|=[2(M_1^2-M_2^2)(\bar{k}^2+
M_i^2)^{1/2}]^{1/2}\langle \Omega_L|a_{i,\bar{k}},
\label{A5}
\end{eqnarray}
as normalized according to
\begin{eqnarray}
\langle k_{L}^{1}|k_{R}^{1}\rangle=1,\qquad \langle k_{L}^{2}|k_{R}^{2}\rangle=1,\qquad \langle k_{L}^{1}|k_{R}^{2}\rangle=0,\qquad \langle k_{L}^{2}|k_{R}^{1}\rangle=0,\qquad |k_{R}^{1}\rangle\langle k_{L}^{1}|+|k_{R}^{2}\rangle\langle k_{L}^{2}|=I.
\label{A6}
\end{eqnarray}
Through use of the commutation relations, in analog to (\ref{N31})  we obtain the on-shell contributions
\begin{eqnarray}
&&\langle \Omega_L|\bar{\phi}(0)|k_{R}^1\rangle \langle k_{L}^1|\bar{\phi}(0)|\Omega_R \rangle
=-\frac{1}{(2\pi)^32(\bar{k}^2+M_1^2)^{1/2}(M_1^2-M_2^2)},
\nonumber \\
&&\langle \Omega_L|\bar{\phi}(0)|k_{R}^2\rangle \langle k_{L}^2|\bar{\phi}(0)|\Omega_R \rangle
=\phantom{-}\frac{1}{(2\pi)^32(\bar{k}^2+M_2^2)^{1/2}(M_1^2-M_2^2)}.
\label{A7}
\end{eqnarray}
Thus just as we found in the PU case, in analog to (\ref{N31}) the relative minus sign in $\bar{D}(x)$ as given in (\ref{M82}) originates in the relative minus sign given in (\ref{A7}), with all on-shell intermediate states having positive norm.

\end{document}